\documentclass[pra,preprint,11pt,superscriptaddress]{revtex4-1} % On govt laptop
\usepackage{amsmath}
\usepackage{amssymb}
\usepackage{graphicx}
\usepackage{color}
\newcommand{\be}[1]{\begin{equation}\label{#1}}
\newcommand{\ee}{\end{equation}}
\newcommand{\bea}[1]{\begin{eqnarray}\label{#1}}
\newcommand{\eea}{\end{eqnarray}}
\newcommand{\no}{\nonumber \\}
\newcommand{\Fig}[1]{Fig.(\ref{#1})}

\newcommand{\Eq}[1]{Eq.(\ref{#1})}
\newcommand{\App}[1]{Appendix~\ref{#1}}
\newcommand{\Sec}[1]{Section~\ref{#1}}
\newcommand{\bsub}{\begin{subequations}}
\newcommand{\esub}{\end{subequations}}

\newcommand{\fracd}[2]{\frac{\displaystyle #1}{\displaystyle #2}}
\newcommand{\om}{\omega}

\def\a0{{\alpha_0}}

\def\da0{{\dot{\alpha}_0}}

\def\myover#1{\myoverDefn#1}
\def\myoverDefn#1#2{\hbox{\space \raise-2mm\hbox{$\textstyle{#1} \atop \scriptstyle{#2}$} }}
\def\om{{\omega}}

\def\G{{\Gamma}}

\def\g{{\gamma}}

\def\a{{\alpha}}

\def\vac{\textrm{vac}}
\def\dag{\dagger}
\def\exia{e^{i\xi_a L}}
\def\exib{e^{i\xi_b L}}
\def\exik{e^{i\xi_k L}}

\def\rp{r_{P}}
\def\rp2{r_{p}^{2}}
\def\Tr{\textrm{Tr}}
\def\eithp{e^{i \theta_p(\omega)}}
\def\emithp{e^{-i \theta_p(\omega)}}
\def\rab{r_{ab}(\omega)}

\newcommand{\half}{\frac{1}{2}}

\newcommand{\ket}[1]{|#1\rangle}
\newcommand{\bra}[1]{\langle #1|}

\begin{document}
%============================================
%\flushleft{\large DRAFT: rr\_losses\_paper\_v3.tex}
%\vspace{-1em}\flushleft{(\textit{v1: original rough draft; v2: references added; v3: (somewhat) reduced appendices})}
%\vspace{-1em}\flushleft{({\textit{\color{red} Conclusion still needed})}
%% _v4.tex PMA version to all atuhors 18July2016
%% _v5.tex AMS edits 21July2016
%% _v6.tex PMA edits and final proof read - still need to check figure captions
%============================================
\title{Photon pair generation in a lossy microring resonator. I. Theory}
%\title{A quantum optical description of losses in ring resonators: \\
%a comparison and extension of the \\
%Raymer \& McKinstrie and PMA \& EEH approaches \\
%for single bus ring resonator, and a two-photon state input}
\author{Paul M. Alsing}
\affiliation{Air Force Research Laboratory, Information Directorate, 525 Brooks Rd, Rome, NY, 13411}
\author{Edwin E. Hach III}
\affiliation{Rochester Institute of Technology, School of Physics and Astronomy, 85 Lomb Memorial Dr., Rochester, NY 14623}
\date{\today}
%\maketitle

\begin{abstract}
We investigate entangled photon pair generation in a lossy microring resonator using an input-output formalism
based on the work of Raymer and McKinstrie  (Phys. Rev. A \textbf{88}, 043819 (2013))
%\cite{Raymer:2013}
%\textit{Quantum input-output theory for optical cavities with arbitrary coupling strength: Application to two-photon wave-packet shaping}, Phys. Rev. A \textbf{88}, 043819 (2013) paper (designated RM)
and Alsing, \textit{et al.} (Phys. Rev. A \textbf{95}, 053828 (2017))
%\cite{Alsing_Hach:2016}
%\textit{A quantum optical description of losses in ring resonators based on field operator transformations}, arxiv:1608.01280  \cite{Alsing_Hach:2016} (designated AH).
that incorporates  circulation factors that account for the multiple round trips  of the fields within the cavity.
We consider the nonlinear processes of spontaneous parametric down conversion and spontaneous four wave mixing,
and we compute the generated biphoton signal-idler state from a single bus microring resonator, along with the generation, coincidence-to-accidental, and heralding efficiency rates.
We compare these generalized results to those obtained by previous
works employing the standard Langevin input-output formalism.
%In addition, we examine the entanglement of the biphoton state as a function of the microring resonator coupling and %internal propagation loss parameters.
\end{abstract}
\maketitle
%==================================
\section{Introduction}\label{sec:intro}
%==================================
Over the last decade, advances in chip-based fabrication have made micron-scale, high quality factor Q
integrated optical microring resonators (mrr) coupled to an external bus ideal sources of entangled photon pair generation, requiring only $\mu$Ws of pump power \cite{Lipson:2010,Sipe:2012a,Popovic:2015,Preble:2015,Tison:2017,Vernon:2017}.
Such high-Q mircoring resonators exhibit nonlinear optical properties allowing for biphoton generation arising from the $\chi^{(2)}$ processes of spontaneous parametric down conversion (SPDC), and the $\chi^{(3)}$ processes of spontaneous four-wave mixing (SFWM). Much theoretical research has been devoted to study the generation of entangled photon pair within cavities and mrr in the weak pump field driving limit \cite{Scholz:2009,Shen_Fan:2009a,Shen_Fan:2009b,Tsang:2011,Camacho:2012,Sipe:2012a,Sipe:2012b,Sipe:2015a,Sipe:2015c},
and more recently in the strong pump field regime \cite{Sipe:2015b} where higher order nonlinear effects such as self-phase and cross-phase modulation become important.

The predominant method of analysis for analyzing a driven cavity or mrr is
the standard Langevin input-output formalism \cite{Collett_Gardiner:1984,Walls_Milburn:1994,Mandel_Wolf:1995,Scully_Zubairy:1997,Orszag:2000} which allows one to express the output field in terms of the intra-cavity and external driving fields. This formalism is valid in the high cavity Q limit, near cavity resonances, but does not adequately address processes throughout the entire free spectral range of the cavity.
In this work we investigate entangled photon pair generation in a microring resonator using a recent input-output formalism based on the work of Raymer and McKinstrie  \cite{Raymer:2013}
%\textit{Quantum input-output theory for optical cavities with arbitrary coupling strength: Application to two-photon wave-packet shaping}, Phys. Rev. A \textbf{88}, 043819 (2013) (designated RM),
and Alsing, \textit{et al.} \cite{Alsing_Hach:2016}
%\textit{A quantum optical description of losses in ring resonators based on field operator transformations}, arxiv:1608.01280  \cite{Alsing_Hach:2016} (designated AH),
that incorporates the circulation factors that account for the multiple round trips  of the fields within the cavity.
We consider biphoton pair generation within the mrr via both SPDC and SFWM, and compute the generated two-photon signal-idler intra-cavity and output state from a single bus (all-through) microring resonator.
We also compute the two-photon generation, coincidence-to-accidental, and heralding efficiency rates.
We compare our results to related calculations \cite{Scholz:2009, Tsang:2011, Sipe:2015a}  obtained using the
standard Langevin input-output formalism.
%In addition, we investigate the entanglement of the biphoton state generated within the mrr as a function of the mrr coupling and internal propagation loss.

This paper is organized as follows.
% Section II
In \Sec{sec:SPDC:SFWM:processes} we derive and solve the equations of motion for the pump, signal and idler fields within a mrr coupled to a single external bus using a combination of the formalism of Raymer and McKinstrie \cite{Raymer:2013}
and Alsing, \textit{et al.} \cite{Alsing_Hach:2016}. We consider the weak, non-depleted pump field limit where higher order nonlinear processes such as self-phase and cross-phase modulations effects are neglected.
We also examine the commutators of the quantum noise fields introduced to account for internal propagation loss which need not commute within the mrr, a phenomena noted by previous authors  Barnett \cite{Barnett:1996} and Agarwal \cite{Agarwal:2014} in their study of circulating cavity fields. In contrast to the standard Langevin approach, we show these commutators, which can be uniquely solved for by requiring the unitarity of the input and output fields, contain pump dependent contributions.
%
% Section III
In \Sec{sec:2photon_state:outside} we compute the output biphoton state, and calculate its generation rate, along with the coincidence to accidental, and heralding efficiency rates.
%
% Section IV
In \Sec{sec:2photon_state:inside_mrr} we compute the biphoton state generated within the mrr, since this is the state most often calculated in the literature and affords the most straightforward comparison. Again, we calculate the biphoton generation-, coincidence to accidental, and heralding efficiency rates. We investigate how the mrr self-coupling (bus-bus, mrr-mrr) and internal propagation loss effects these rates.
%
% Section V
%In \Sec{sec:squeezed:state} we examine the form of the output squeezed vacuum state from a Heisenberg operator perspective. This allows us to form the operator that generates the output squeezed  state, as well as the unitary operator that evolves the external input operators to the output operators. We further examine the entanglement of the output squeezed state employing the log negativity and its dependence on the mrr coupling and internal propagation loss.
%
% Section VI
In \Sec{sec:summary:discussion} we summarize our results and indicate avenues for future research.
%
%In the \App{app}
%\App{app:derivation:single:bus}
%we review the standard Langevin input-output formalism, and the modifications introduced by Raymer and McKinstrie, and Alsing, \textit{et al.} to account the circulating contributions of the intracavity fields.
%
In the \App{app} %\App{app:highQ:limit:G:H}
we examine the our expressions for the output fields, and for rates derived from them, in the high cavity Q limit where the standard Langevin input-output formalism is valid, and compare with prior works in the literature.

\section{SPDC and SPFM processes inside a (single bus) microring resonator}\label{sec:SPDC:SFWM:processes}
\subsection{Preliminaries}\label{subsec:SPDC:SFWM:preliminaries}
\begin{figure}[h]
%\begin{tabular}{cc}
%\includegraphics[width=3.0in,height=2.5in]{rr_loss_all_thru_RR_a_c_aint} &
%\includegraphics[width=3.0in,height=2.5in]{rr_losses_all_thru_RR_a_c} % Original Figure name
\includegraphics[width=3.0in,height=2.5in]{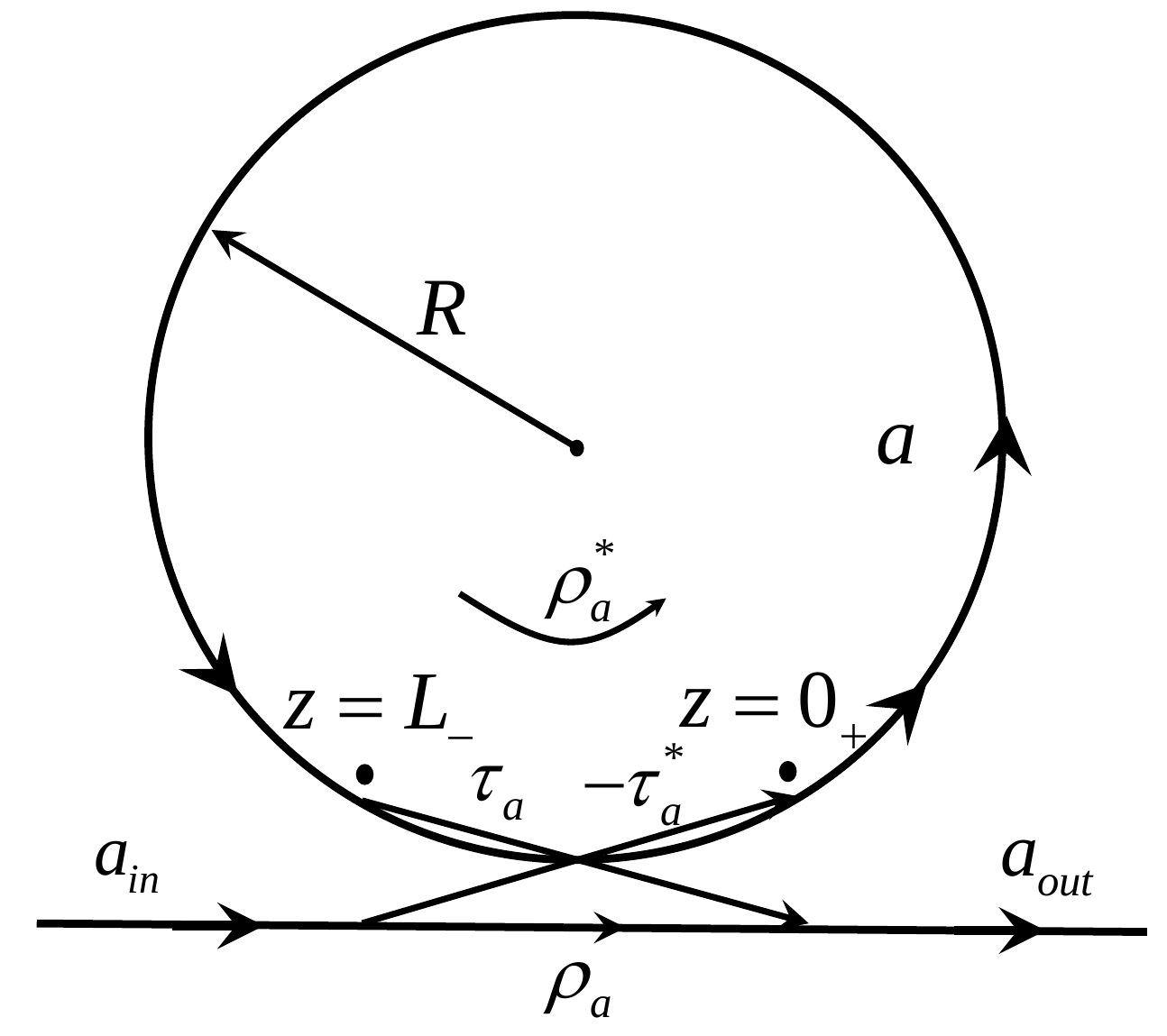}%{fig1_ppg_single_bus_rr_RM_notation}
%\end{tabular}
\caption{A single bus (all-through) microring resonator (mrr) of length $L=2\pi R$ with intracavity field $a$, coupled to a waveguide bus with input field $a_{in}$ and output field $a_{out}$. $\rho_a\, \tau_a$ are the beam splitter like self-coupling and cross-coupling strengths, respectively, of the bus to the mrr such that $|\rho_a|^2 + |\tau_a|^2 =1$. $z=0_+$ is the point just inside the mrr which cross-couples to the input field $a_{in}$, and $z=L_-$ is the point after one round trip in the mrr that cross-couples to the output field $a_{out}$.
}\label{fig:single_bus_rr}
\end{figure}

In this section we examine the nonlinear processes of spontaneous parametric down conversion (SPDC) and spontaneous four wave mixing (SFWM) inside a
single bus microring resonator (mrr) of length $L=2\pi R$, as illustrated in \Fig{fig:single_bus_rr}.
Here, $a$ is the intracavity field which is coupled to a waveguide bus with input field $a_{in}$ and output field $a_{out}$.
%==================================
%==================================
The parameters $\rho_a,\,\tau_a$ are the beam splitter like self-coupling and cross-coupling strengths, respectively, of the bus to the mrr such that $|\rho_a|^2 + |\tau_a|^2 =1$. $z=0_+$ is the point just inside the mrr which cross-couples to the input field $a_{in}$, and $z=L_-$ is the point after on round trip in the mrr that cross-couples to the output field $a_{out}$.
%=================
% Orig from _v2
%================
%In \App{app:derivation:single:bus} we derive the equation of motion for an intracavity field $a$ inside the mrr
%based on the work of Raymer and McKintrie \cite{Raymer:2013} and Alsing \textit{et al} \cite{Alsing_Hach:2016},
%taking into account both coupling and internal propagation loss.

In the work of Raymer and McKinstrie \cite{Raymer:2013} (abbreviated as RM) the cavity field $a$ satisfies a
traveling-wave Maxwell ODE in the absence of internal propagation loss  given by
\be{RM:1:v3}
(\partial_t + v_a\,\partial_z)\, a(z,t) = \alpha_{polz}\,P(z,t),
\ee
where $a(z,t)$ is the ring resonator cavity field (in the time domain), $v_a$ is the group velocity, $P(z,t)$ is the polarization and $\alpha_{polz}$ is a coupling constant. The carrier wave frequency has been factored out so that all frequencies are relative to the optical carrier frequency. The input coupling and periodicity of the cavity is captured by the boundary conditions
\bsub
\bea{RM:2:v3}
 a(0_+,t) &=& \rho_a\,a(L_-,t) + \tau_a\, a_{in}(t), \\
 a_{out}(t)  &=& \tau_a\,a(L_-,t) - \rho_a\, a_{in}(t), \label{RM:3:v3}
\eea
\esub
where we have taken the  beam splitter like  self-coupling $\rho_a$ (buss-bus, mrr-mrr),
and cross-coupling $\tau_a$ (bus-mrr) real for simplicity
%such that $|\rho_a|^2 + |\tau_a|^2 =1$,
and the minus sign in \Eq{RM:3:v3} accounts for the $\pi$ change in phase arising from the "reflection" of the input field off the higher index of refraction mrr to the output (bus) field.
%$z=0_+$ is the point just inside the cavity through which the input field couples with cross coupling strength $\tau_a$,
%while $z=L_-$ is the point inside the mrr just before the cavity field exits to the output field.
The input and output fields satisfy the free field commutators
 \be{RM:5:34}
 [ a_{in}(t), a_{in}^\dagger(t')] = \delta(t-t') = [ a_{out}(t), a_{out}^\dagger(t')].
\ee

The output field $a_{out}(\om)$ is easily solved from
\Eq{RM:1:v3}, \Eq{RM:2:v3} and \Eq{RM:3:v3} in the Fourier domain
%\Eq{RM:10} and \Eq{RM:13}
yielding the unimodular transfer function $G_{out, in}(\om)$ defined by
\be{RM:16:17:v3}
 a_{out}(\om) \equiv  G_{out, in}(\om) \, a_{in}(\om), \qquad G_{out, in}(\om) =  e^{i\om T_a}\,\left[\frac{1-\rho_a\,e^{-i\om T_a}}{1-\rho_a\,e^{i\om T_a}}\right],
 \qquad |G_{out, in}(\om)| = 1.
\ee
Note that in the classical case (see e.g. Yariv\cite{Yariv:2000}, and Rabus \cite{Rabus:2007})
one obtains the result with phenomenological  loss factor $0\le\alpha_a\le1$
\be{AH:63:v3}
a_{out}(\om) \equiv G^{(\alpha)}_{out, in}(\om)\,a_{in}(\om) =
\left(
\frac{\alpha_a\,e^{i\theta_a} - \rho_a}{1-\rho_a^*\,\alpha_a\,e^{i\theta_a}}
\right)\,a_{in}(\om), \qquad  |G^{(\alpha)}_{out, in}(\om)|\le 1,
\ee
which is the same coefficient that appears in the quantum derivation with loss
(see Eq.(13g) in Alsing \textit{et al.} \cite{Alsing_Hach:2016} (abbreviated as AH)
with $\rho_a\rightarrow\tau_a$ real) and $\theta_a = \om\,T_a$.
The lossless case corresponds to $\alpha_a\rightarrow 1$.

%======================
For the quantum derivation including internal propagation loss (generalizing the the lossless mrr considerations begun in \cite{Hach:2014}), AH \cite{Alsing_Hach:2016} used an expression by Loudon \cite{,Loudon:1997,Loudon:2000}  for the attenuation loss of a traveling wave, modeled from a continuous set of beams splitters acting as scattering centers
\cite{,Loudon:1997,Loudon:2000},
\be{aL:body:v3}
a(L_-,\om) = e^{i\xi_a(\omega)L} \, a(0_+,\omega) + i \sqrt{\G_a(\omega)}\,\int_{0}^{L} dz \,e^{i\xi_a(\omega)(L-z)}\,s(z,\omega),
\ee
where $\exia \equiv \alpha_a\,e^{i\,\theta_a}$ with $\alpha_a=e^{-\half\,(\G_a/v_a)\,L}$, $\theta_a = (\om\,n(\om)/c)\,L=\omega\,T_a$  and $\G_a$ incorporates both coupling and internal propagation losses.
Here, $s(z,\omega)$ are the noise scattering operators that give rise to the internal loss and satisfy $[s(z,\omega),s^\dag(z',\omega')] = \delta(z-z')\,\delta(\om-\om')$.
AH explicitly showed that $a(L_-,\om)$ in \Eq{aL:body:v3} satisfied $[a(L_-,\om),a^\dag(L_-,\om')] = \delta(\om-\om')$. By tracking the infinite number of round trip circulations of the cavity field in the single bus mrr, AH derive the expression (with $\tau\rightarrow\rho_a$ and $\kappa\rightarrow\tau_a$ in \cite{Alsing_Hach:2016})
\be{aout:AH}
a_{out}(\om)=
\left(
\frac{\rho_a-\alpha_a\,e^{i\theta_a}}{1-\rho^*_a\,\alpha_a\,e^{i\theta_a}}
\right)\,a_{in}(\om)
-i |\tau_a|^2\,\sqrt{\G_a}\sum_{n=0}^\infty (\rho_a)^n\,
%\Int{(n+1)}
\int_{0}^{(n+1) L}dz\,e^{i\xi_a(\omega)[(n+1) L - z]} \hat{s}(z,\omega).
\ee
AH show by explicit calculation that the output field satisfies $[a_{out}(\om), a_{out}^\dag(\om)] = \delta(\om-\om')$.
In general, this allows one to write
\be{aout:AH:v3}
a_{out}(\om) = G_{out,in}(\om)\,a_{in} + H_{out,in}(\om)\,f_a(\om), \quad |H_{out,in}(\om)| = \sqrt{1 - |G_{out,in}(\om)|^2},
\ee
which defines the quantum noise operator $f_a(\om)$ from the unitary requirement of the preservation of the free field output commutator.
%In \App{app:derivation:single:bus:highQ:limit}
In the \App{app} we examine  $G_{out,in}(\om)$ and $H_{out,in}(\om)$
in the high cavity Q limit, where the standard Langevin input-output formalism is valid.

\subsection{Derivation of output operators from input and noise operators}\label{subsec:deriv:output:operators}
For the consideration of nonlinear biphoton pair generation,
we now consider three intracavity fields circulating within the mrr: the signal field $a(z,t)$, the idler field $b(z,t)$ and the pump field $c(z,t)$.
As in the previous section, we work in the interaction picture where the carrier frequencies $\om_d$ for $d\in\{a, b, c\}$ have been removed,
so that the fields are slowly varying in time.
In the interaction picture the nonlinear Hamiltonian for these processes are taken to be
\bsub
\bea{H:SPDC:SFWM}
 \mathcal{H}^{NL}_{spdc} &=&  g_{spdc}\,\left( c\,a^\dag\,b^\dag + h.a. \right), \quad\quad\qquad \om_c = \om_a + \om_b,\\
 \mathcal{H}^{NL}_{sfwm} &=&  g_{sfwm}\,\left( c^2\,a^\dag\,b^\dag + h.a. \right), \qquad 2\,\om_c = \om_a + \om_b.
\eea
\esub
Each field $d(z,t)$ for $d\in\{a,b,c\}$ satisfies the equation of motion and input-output boundary conditions
\bsub
\bea{d:EoM:BCs}
(\partial_t + v_d\,\partial_z)\, d(z,t) &=& -i\,[d(z,t), \mathcal{H}^{NL}] -\frac{\g'_d}{2}\,d(z,t) + \alpha_{polz}\,F_d(z,t), \\
d(0_+,t) &=& \rho_d\,d(L_-,t) + \tau_d\, d_{in}(t),  \label{d:BC:in}\\
d_{out}(t)  &=& \tau_d\,d(L_-,t) - \rho_d\, d_{in}(t), \label{d:BC:out}
\eea
\esub
where we have included the internal mrr propagation loss given by the rate  $\gamma'_d$.
We also allow for different group velocities $v_d(\om) = c/n_d(\om)$ for each mode $d$ leading to different round trip times $T_d = L/v_d$ for $k\in\{a,b,c\}$.
Junction coupling losses between the ring resonator and the bus are taken into account by later defining the
self-coupling loss $\gamma_d$  via $\rho_d\equiv e^{-\gamma_d\,T_d/2}$ \cite{Raymer:2013}.

In the above $F_d(z,t)$ are the noise operators inside the ring resonator and $\alpha_{polz}$ is a coupling constant of the internal modes $a,b$ to the polarization field, giving rise to internal loss (see RM \cite{Raymer:2013}). While the noise operators could be derived directly as in AH \cite{Alsing_Hach:2016} by tracking the multiple round trips of each field $d$ through the mrr, here we have opted for the Langevin-based approach indicated (but not explored) in RM  \cite{Raymer:2013}. Here, we differ from RM by not explicitly indicating the value of the commutation relations for the noise operators $F_d(z,t)$, preferring instead to compute their values later
by the causality condition that the output fields $ d_{out}$ of the above coupled set of equations satisfy the free field canonical commutation relations, given that the input fields $d_{in}$ do. The particular value of the commutators are important when we compute the reduced density matrix $\rho_{ab}$ for the two-photon output signal-idler state.
For now the noise operators $F_d$ are simply carried along through the computation.

The above equations are most easily solved in the frequency domain
\footnote{Note that $\om$ is the offset from the central pump frequency $\om_p$ so that $b^\dag(\om) \equiv [b(-\om)]^\dag$ (see \cite{Orszag:2000}).}
(using $d(z,\om) = \int_{-\infty}^{\infty} dt\, d(z,t)\,e^{i\om\,t}$, for $d\in\{a,b,c\}$, and
$f_d(z,\om) = \int_{-\infty}^{\infty} dt\, F_d(z,t)\,e^{i\om\,t}$).
Here the interaction Hamiltonians are given by
%%***(\textit{\color{red} double check value of $g_{spdc}(\om)$})
\bsub
\bea{H:SPDC:SFWM:om}
 \mathcal{H}^{NL}_{spdc} &=&  \int_{-\infty}^{\infty}\,\frac{d\om}{2\pi}\,g_{spdc}(\om)\,\left( c(z,\om)\,a^\dag(z,\om)\,b^\dag(z,\om) + h.a. \right),
 \quad g_{spdc}(\om) = \frac{3\,(\hbar\,\om_c)^{3/2}\,\chi^{(2)}}{4\epsilon_0\,\bar{n}^4\,V_{ring}}, \qquad\\
 \mathcal{H}^{NL}_{sfwm} &=&  \int_{-\infty}^{\infty}\,\frac{d\om}{2\pi}\,g_{sfwm}(\om)\,\left( c^2(z,\om)\,a^\dag(z,\om)\,b^\dag(z,\om) + h.a. \right),
 \quad g_{sfwm}(\om) = \frac{3(\hbar\,\om_c)^2\,\chi^{(3)}}{4\epsilon_0\,\bar{n}^4\,V_{ring}}, \qquad.
\eea
\esub
where the value of $g_{spdc}(\om)$ and $g_{sfwm}(\om)$ \cite{Sipe:2012b,Sipe:2015a} depend on the volume $V_{ring}$ of the ring mode,
and the nonlinear susceptibilities $\chi^{(2)}$ and $\chi^{(3)}$ are accessed uniformly in the ring. Here $\bar{n}$ is the average index of refraction of the ring (assumed constant) and $\epsilon_0$ is the permittivity of free space.

In the undepleted pump approximation, employed here, the equation of motion for the pump mode $c$ satisfies \Eq{RM:1:v3}
(with $a\rightarrow c$ and $P(z,t)=0$), and Hamiltonian terms such as $-i\,g_{spdc}\,a\,b$ and $-i\,g_{sfwm}\,c^\dag\,a\,b$ are considered small, and hence dropped along with the noise term $f_c$ \footnote{This corresponds to dropping higher order terms describing self-phase and cross-phase pump modulations terms, see \cite{Sipe:2015a, Sipe:2015b}}. This equation is then classical, and the value of the lossless pump inside the ring becomes
%(see \Eq{RM:13})
\be{c:mrr}
  \langle c(0_+,\om) \rangle = \frac{\tau_c}{1-\rho_c\,e^{i\om T_c}}\, \langle c_{in}(\om)\rangle, \quad
  \langle c(L_-,\om) \rangle = \frac{\tau_c\,e^{i\om T_c}}{1-\rho_c\,e^{i\om T_c}}\, \langle c_{in}(\om)\rangle.
\ee
where the angled brackets indicate that we are dealing with a c-number field value.
Outside the ring, the pump field is given by
%(see \Eq{RM:16:17})
\be{c:out}
 \langle c_{out}(\om)\rangle \equiv  G_{c_{out}\,c_{in}}(\om) \,\langle c_{in}(\om)\rangle, \qquad G_{c_{out}\,c_{in}}(\om) =  e^{i\om T_c}\,\left[\frac{1-\rho_c\,e^{-i\om T_c}}{1-\rho_c\,e^{i\om T_c}}\right].
\ee
Therefore, in the Hamiltonian we replace the operator $c_p$ by $\langle c(z,\om) \rangle$  and write
\bsub
\bea{H:NL}
 \mathcal{H}^{NL} &=&  \int_{-\infty}^{\infty}\,\frac{d\om}{2\pi}\,g(\om)\,
 \left( \alpha_p(z,\om)\, a^\dag(z,\om)\,b^\dag(z,\om) + \alpha^*_p(z,\om)\,a(z,\om)\,b(z,\om) \right), \\
g(\om) &=& g_{spcd}(\om), \quad\;\; \alpha_p(z,\om) =  \langle c(z,\om) \rangle \;\; \textrm{for SPDC}, \\
g(\om) &=& g_{sfwm}(\om), \quad \alpha_p(z,\om) =  \langle c^2(z,\om) \rangle \;\; \textrm{for SFWM}.
\eea
\esub
Thus, for both nonlinear processes the signal and idler modes satisfy the equation of motion  in the frequency domain
\bsub
\bea{a:b:EoM}
(-i\,\om + v_a\,\partial_z)\, a(z,\om) &=&  -i\,g\,\alpha_p(z,\om)\,b^\dag(z,\om) -\frac{\g'_a}{2}\,a(z,\om) + \alpha_{polz}\,f_a(z,\om), \label{a:EoM}\\
(-i\,\om + v_b\,\partial_z)\, b^\dag(z,\om) &=&   i\,g\,\alpha^*_p(z,\om)\,a(z,\om) -\frac{\g'_b}{2}\,b^\dag(z,\om) + \alpha_{polz}\,f_b(z,\om). \label{b:EoM}
\eea
\esub
\Eq{a:b:EoM} has the formal solution
\be{soln:a}
a(L_-,\om) = a(0_+,t)\,\exia + \int_0^L dz'\, \left( (-i g \alpha_P/v_a)\, b^\dag(z',\om) + (\alpha_{polz}/v_a)\,\tilde{f}_a(z',\om) \right)\,e^{i\,\xi_a (L-z')}
\ee
where $\xi_a  = (\om + i\,\gamma'_a/2)/v_a$ so that $i\,\xi_a L = (i\om -\,\gamma'_a/2)\,T_a$.
For fast molecular damping we approximate the last term by setting $z'\rightarrow L$ (the upper limit of the integral)
and factoring out $\alpha_p(L_-,\om)\,b^\dag(L_-,\om)$ from the integral.
The remaining integral yields
$(-i\,g\,\alpha_p(L_-,\om)/v_a)\,\int_0^L dz' \,e^{i\,\xi_a (L-z')}$ $= (-i g \alpha_p(L_-,\om)/v_a)(1-\exia)/(-i\xi_a)\rightarrow -i g\,\alpha_p(L_-,\om)\,T_a \equiv -i\,r_a(\om)$ defining the dimensionless pump parameter $r_a(\om) = g\,\alpha_p(\om)\,T_a$.
Thus, we write
\Eq{soln:a} as
\bsub
\be{a:L}
a(L_-,\om) = a(0_+,t)\,\exia -i\,r_a(\om)\,b^\dag(L_-,t) + f_a(\om), \quad f_a(\om)\equiv(\alpha_{polz}/v_a) \int_0^L dz'\,\tilde{f}_a(z',\om)\,e^{i\,\xi_a (L-z')}. \qquad\qquad
\ee
%where we have defined
%$f_a(\om)=(\alpha_{polz}/v_a) \int_0^L dz'\,\tilde{f}_a(z',\om)\,e^{i\,\xi_a (L-z')}$.
Similarly, \Eq{b:EoM} yields
\be{b:L}
b^\dag(L_-,\om) = b^\dag(0_+,t)\,\exib +i\,r_b(\om)\,a(L_-,t) + f^\dag_b(\om), \; f^\dag_b(\om)=(\alpha_{polz}/v_a) \int_0^L dz'\,\tilde{f}^\dag_b(z',\om)\,e^{i\,\xi_b (L-z')}, \qquad\qquad
\ee
\esub
where we have defined $f_a(\om)$ and $f^\dag_b(\om)$
and used the notation
$r_a(\om) \equiv g\,\alpha_p(L_-,\om)\,T_a$ and $r_b(\om) \equiv g\,\alpha^*_p(L_-,\om)\,T_b$.
%$f_b(\om)=(\alpha_{polz}/v_a) \int_0^L dz'\,\tilde{f}_b(z',\om)\,e^{i\,\xi_b (L-z')}$.
We can therefore put equations of motion and boundary conditions for the signal $a$ and idler $b$ modes in matrix form as
\bsub
\bea{eqns:matrix:form}
 M \, \vec{a}(L_-,\om) &=& P_\xi\, \vec{a}(0_+,\om) + \vec{f}(\om), \label{eqns:matrix:form:eom}\\
 \vec{a}(0_+,\om) &=& T_\rho \, \vec{a}(L_-,\om) + X_\tau \, \vec{a}_{in}(\om), \label{eqns:matrix:form:BC:in}\\
 \vec{a}_{out}(\om) &=& X_\tau \, \vec{a}(L_-,\om) - T_\rho  \, \vec{a}_{in}(\om), \label{eqns:matrix:form:BD:out}
\eea
\esub
where we have defined
\be{defns:matrices}
M = \left(
      \begin{array}{cc}
        1 & i\,r_a \\
        -i\,r_b & 1 \\
      \end{array}
    \right), \;
P_\xi = \left(
      \begin{array}{cc}
        \exia & 0 \\
        0 & \exib \\
      \end{array}
    \right), \;
T_\rho = \left(
      \begin{array}{cc}
        \rho_a & 0 \\
        0 & \rho_b \\
      \end{array}
    \right), \;
X_\tau = \left(
      \begin{array}{cc}
        \tau_a & 0 \\
        0 & \tau_b \\
      \end{array}
    \right),
\ee
 and
\be{defns:vectors}
 \vec{a}(\om) =
 \left(
   \begin{array}{c}
     a(\om) \\
     b^\dag(\om) \\
   \end{array}
 \right), \quad
 \vec{a}_{in}(\om) =
 \left(
   \begin{array}{c}
     a_{in}(\om) \\
     b_{in}^\dag(\om) \\
   \end{array}
 \right), \quad
 \vec{a}_{out}(\om) =
 \left(
   \begin{array}{c}
     a_{out}(\om) \\
     b_{out}^\dag(\om) \\
   \end{array}
 \right), \quad
\vec{f}(\om) =
 \left(
   \begin{array}{c}
     f_a(\om) \\
     f_b^\dag(\om) \\
   \end{array}
 \right). \qquad\qquad
\ee
Here $T_\rho$ represents the through coupling `reflection' from input bus off the ring resonator to output bus (and the self-coupling within the mrr),
while $X_\tau$ represents the cross coupling `transmission' between the bus and the ring resonator. The term $P_\xi$ represents the roundtrip phase accumulation and intrinsic loss within the ring resonator, and we define $\exik \equiv \alpha_k \, e^{i\theta_k}$ with $\alpha_k = e^{-\gamma'_k\,T_k/2}$ and $\theta_k = \om\,T_k$,
$r_a = g\,\alpha_p\,T_a$, and $r_b = g\,\alpha^*_p\,T_b$.

A substitution of $\vec{a}(0_+,\om)$ from \Eq{eqns:matrix:form:BC:in} into the right hand side of \Eq{eqns:matrix:form:eom} allows for the solution of the intracavity field (just before exit) $\vec{a}(L_-,\om)$ in terms of $\vec{a}_{in}(\om)$ and $\vec{f}(\om)$. A subsequent substitution of this solution for $\vec{a}(L_-,\om)$ into the right hand side of \Eq{eqns:matrix:form:BD:out} produces the desired output field $\vec{a}_{out}(\om)$ in terms of the input field  $\vec{a}_{in}(\om)$ and noise operators $\vec{f}(\om)$.
After some lengthy but straightforward algebra, the output fields can be expressed in terms of the input fields as
\be{aout:matrix:form}
\vec{a}_{out}(\om) = G(\om) \, \vec{a}_{in}(\om) + H(\om)\, \vec{f}(\om),
\ee
where
\bsub
\bea{Gom:v2}
 G(\om) &=& \left[ X_\tau  (M - P_\xi\,T_\rho)^{-1}\right] P_\xi\, X_\tau - T_\rho
\equiv
\left(
  \begin{array}{cc}
    G_{aa}(\om) & G_{ab}(\om) \\
    G_{ba}(\om) & G_{bb}(\om) \\
  \end{array}
\right), \label{Gom:v2:matrixform}\\
&\equiv& \frac{1}{D}
\left(
  \begin{array}{cc}
    (\exia - \rho_a )\,(1 - \rho_b\,\exib) + r_a\,r_b\,\rho_a  &   -i\,r_a\,\tau_a\,\tau_b\,\exib \\
    i\,r_b\,\tau_b\,\tau_a\,\exia       &  (\exib - \rho_b)\,(1 - \rho_a\,\exia) + r_a\,r_b\,\rho_b \\
  \end{array}
\right), \qquad \label{Gom:v2:matrix:elements}
\eea
\esub
%=========================
%with
%\bsub
%\bea{defns:S_D_dab:v2}
%S_k &=& \frac{1}{1-\rho_k\,\exik} = \sum_{n=0}^{\infty}\, \left(\rho_k\,\exik\right)^n \equiv \sum_{n=0}^{\infty}\, \left(\rho_k\,\alpha_k\,e^{i\,\theta_k}\right)^n,\;
%\alpha_k = e^{-\gamma'_k/2\,T_k}, \; \theta_k = \om\,T_k, \qquad \\
%%
%S_k-1 &=& \rho_k\,\exik\,S_k, \qquad
%\tilde{D}_{ab} = 1 - \rp2\,S_a\,S_b, \qquad \rp2 \equiv r_a\,r_b = (g\alpha_P T_a)\,(g\alpha^*_P T_b),\qquad
%\eea
%\esub
%=========================
with
\bsub
\bea{defns:S_D_dab:v2}
D &=& (1 - \rho_a\,\exia)\,(1 - \rho_b\,\exib) - r_a\,r_b, \qquad r_a = g\alpha_P T_a\, \quad \,r_b =g\alpha^*_P T_b,\\
%S_k &=& \frac{1}{1-\rho_k\,\exik} = \sum_{n=0}^{\infty}\, \left(\rho_k\,\exik\right)^n \equiv \sum_{n=0}^{\infty}\, \left(\rho_k\,\alpha_k\,e^{i\,\theta_k}\right)^n,\;
\alpha_k &=& e^{-\gamma'_k/2\,T_k}, \quad \theta_k = \om\,T_k, \quad \textrm{for}\quad k\in\{a,b\},
%%
%S_k-1 &=& \rho_k\,\exik\,S_k, \qquad
%\tilde{D}_{ab} = 1 - \rp2\,S_a\,S_b, \qquad \rp2 \equiv r_a\,r_b = (g\alpha_P T_a)\,(g\alpha^*_P T_b),\qquad
\eea
\esub
%for $k\in\{a,b\}$
and
%==========
% H better
%==========
\be{Hom:v2}
H(\om) = X_\tau  (M - P_\xi\,T_\rho)^{-1}
\equiv
\left(
  \begin{array}{cc}
    H_{aa}(\om) & H_{ab}(\om) \\
    H_{ba}(\om) & H_{bb}(\om) \\
  \end{array}
\right)
= \frac{1}{D} \,
\left(
  \begin{array}{cc}
    \tau_a\,(1-\rho_b\exib)  & -i\,r_a\,\tau_a \\
     i\,r_b\,\tau_b          & \tau_b\,(1-\rho_a\exia)
  \end{array}
\right). \qquad
\ee
%==========================================
Note that to lowest order in $|g\alpha_p|$, we have $1/D\approx S_a\,S_b$
%where $S_k$
where
$S_k = \frac{1}{1-\rho_k\,\exik} = \sum_{n=0}^{\infty}\, \left(\rho_k\,\exik\right)^n \equiv \sum_{n=0}^{\infty}\, \left(\rho_k\,\alpha_k\,e^{i\,\theta_k}\right)^n$
for $k\in\{a,b\}$
are the geometric series  factors resulting from the round trip circulations of the internal fields $k\in\{a,b\}$ inside the ring resonator. For typical ring resonator of radius $R=20 \mu$m and pump laser power of 1mW ($\chi^{(2)}\sim 2\times 10^{-12}$ m/V, $\alpha_p\sim 10^3$ V/m,) and round trip times of $T_k\sim$ 1 ps, we have $r_p\sim 10^{-5}$ \cite{Schneeloch:2017}.
 %A bit of straightforward algebra reveals that
% $G(\om) = X_\tau  (M - P_\xi\,T_\rho)^{-1}\, P_\xi\, X_\tau - T_\rho = X_\tau \, M^{-1}\, P_\xi \, (I - T_\rho\,M^{-1}\,P_\xi)^{-1} \, X_\tau - T_\rho$
% and
% $H(\om) = X_\tau  (M - P_\xi\,T_\rho)^{-1} = X_\tau \left[ \, M^{-1}\, P_\xi \, (I - T_\rho\,M^{-1}\,P_\xi)^{-1} \, T_\rho + I \right] \,M^{-1}$ so that the formal matrix expressions in this section agree with those in the previous section. The difference between the two is the use of the geometric series $S_k$ in this section versus the power dependent series $\SDk$ in the previous section.
 A comparison of the matrix forms of $G(\om)$ in \Eq{Gom:v2:matrixform} and $H(\om)$ in \Eq{Hom:v2} reveals the useful relationship
\be{G:H}
 G(\om) = H(\om)\,P_{\xi}(\om)\,X_{\tau}(\om) - T_{\rho}(\om).
\ee
In the \App{app}
%\App{app:highQ:limit:G:H}
we examine $G(\om)$ and $H(\om)$ in the high cavity Q limit defined by
$\rho_k\equiv e^{-\gamma_k\,T_k/2}\rightarrow 1,\; \om\,T_k\ll 1$
%$\rho_k\rightarrow 1, \om\,T_k\ll 1$,
where the standard Langevin approximation \cite{Walls_Milburn:1994,Orszag:2000} is valid, and compare our results with recent related work \cite{Tsang:2011} using the later formulation.

\subsection{Commutators of the noise operators}\label{subsec:comm:noise:operators}
\subsubsection{Linear equations determined by causality}\label{subsubsec:linear:equations}
The commutation relations between the noise operators are fundamentally determined by the canonical commutators of the free input and output fields.
Given that the input fields satisfy $[a_{in}(\om), a^\dag_{in}(\om')] = [b_{in}(\om), b^\dag_{in}(\om')] = \delta(\om-\om')$, and that they each commute with the
noise operators $f_{a}(\om),\,f_{b}(\om)$ (via causality), one must also have that
$[a_{out}(\om), a^\dag_{out}(\om')] = [b_{out}(\om), b^\dag_{out}(\om')] = \delta(\om-\om')$.
Using \Eq{G:H}, this unitary requirement determines the set of linear equations
\bsub
\bea{comms:spdc}
{} [a_{out}(\om), a^\dag_{out}(\om')]&=& \delta(\om-\om') \Rightarrow  |H_{aa}|^2\, C_{aa} - |H_{ab}|^2\,C_{bb} + 2 \textrm{Re}( H_{aa}\,H^*_{ab}\,D_{ab})
= 1 - (|G_{aa}|^2 - |G_{ab}|^2), \qquad \label{comms:spdc:aadag}\\
{} [b_{out}(\om), b^\dag_{out}(\om')] &=& \delta(\om-\om') \Rightarrow  -|H_{ba}|^2\, C_{aa} + |H_{bb}|^2\,C_{bb} + 2 \textrm{Re}( H_{ba}\,H^*_{bb}\,D_{ab})
= 1 - (|G_{bb}|^2 - |G_{ba}|^2), \label{comms:spdc:bbdag} \qquad\quad\\
{} [a_{out}(\om), b_{out}(\om')] &=& 0 \Rightarrow  H_{aa} \, H^*_{ba} \, C_{aa} - H_{ab}\,H^*_{bb}\,C_{bb} +  H_{aa}\,H^*_{bb}\,D_{ab} + H_{ab}\,H^*_{ba}\,D^*_{ab}
= G_{ab}\,G^*_{bb} - G_{aa}\,G^*_{ba}, \qquad \label{comms:spdc:ab}\\
{} [a_{out}(\om), b^\dag_{out}(\om')] &=& 0 \Rightarrow \textrm{det}(H) \label{comms:spdc:abdag}\, C_{ab} = 0,
\eea
\esub
for the four constants $C_{aa}$, $C_{bb}$, $C_{ab}$, $D_{ab}$ defined by the commutation relations
\bsub
\bea{}
{} [f_{a}(\om), f^\dag_{a}(\om')] = C_{aa}\,\delta(\om-\om'),  \quad [f_{b}(\om), f^\dag_{b}(\om')] = C_{bb}\,\delta(\om-\om'), \\
{} [f_{a}(\om), f^\dag_{b}(\om')] = C_{ab}\,\delta(\om-\om'),  \quad [f_{a}(\om), f_{b}(\om')] = D_{ab}\,\delta(\om-\om').
\eea
\esub
 Since $\textrm{det}(H)\ne 0$, \Eq{comms:spdc:abdag}
 reveals that $C_{ab}=0$. The first three equations are four equations in the four (real) unknowns
 $C_{aa}$, $C_{bb}$, $\textrm{Re}(D_{ab})$, $\textrm{Im}(D_{ab})$ which therefore have a unique solution.
 Note that in the standard Langevin approach \cite{Walls_Milburn:1994, Orszag:2000} one assumes the canonical values $C_{aa}=C_{bb}=1$ and $C_{ab}=D_{ab}=0$. But the standard input-output formalism (here, valid near resonances of the ring resonator) was explicitly constructed so these canonical values identically satisfy the above set of linear equations (see for example the $G$ and $H$ matrices used in Tsang \cite{Tsang:2011} and Shapiro \cite{Shapiro:2015}). These special commutator values are not necessarily valid in general, and in particular $D_{ab}\ne0$ as pointed out in the works of Barnett \cite{Barnett:1996} and Agarwal \cite{Agarwal:2014}. In general, the values of  $C_{aa}$, $C_{bb}$, $D_{ab}$ must be computed from \Eq{comms:spdc:aadag}, \Eq{comms:spdc:bbdag} and \Eq{comms:spdc:ab}.
 The values of these commutators are not only important for the self consistency of the theory, but are also relevant when one computes the accidental singles rate upon the loss of either the generated signal or idler photon due to noise in the ring resonator. However, the values of these commutators do not affect the two-photon portion of the total state (see next section) where neither a signal nor an idler photon is absorbed within the mrr.

 \subsubsection{Exact solution of the commutator equations}\label{subsubsec:soln:comm:equations}
 Using the expressions in \Eq{Gom:v2:matrixform} for $G(\om)$ and \Eq{Hom:v2} for $H(\om)$ a long but straightforward calculation results in the following simple \textit{exact} solutions for the commutator equations \Eq{comms:spdc:aadag}-\Eq{comms:spdc:ab}
 \bsub
 \bea{solns:comms}
 C_{kk}(\om) &=& 1 - \alpha_k^2 - |r_k|^2 = 1 -  e^{-\gamma'_k T_k} - |g\alpha_p\,T_k|^2  \myover{{\longrightarrow} {\textrm{high Q}}} \gamma'_k\,T_k - |g\alpha_p\,T_k|^2, \quad k\in\{a,b\}, \qquad\\
 D_{ab} &=& i\, (r^*_b - r_a) = i\,g\,\alpha_p\,(T_b - T_a), \label{solns:comms:Dab}
 \eea
 \esub
where  for $C_{aa}(\om)$ and $C_{bb}(\om)$ we have also indicated their values in the high cavity Q limit. We note that $C_{kk}(\om)$ for $k\in\{a,b\},$ contains a power dependent correction $|r_k|^2 = |g\,\alpha_p\,T_k|^2$ of higher order than the leading order term $1 - \alpha_k^2$ which approaches $\gamma'_k\,T_k$ is the high Q limit.
%
%Let us further redefine the noise operators as
%$\tilde{f}'_k(\om) \equiv (T_k)^{-1/2}\tilde{f}_k(\om) = [(1-\alpha_k^2)\,T_k]^{-1/2}\,f_k(\om)$
%over those discussed in the previous section on the  high Q limit, such that
%$C_{kk}(\om) \equiv T_k\,\tilde{C}_{kk} = [(1-\alpha_k^2)\,T_k]\tilde{C}'_{kk}(\om)$.
%Then, in the high Q limit, to first order in the pump parameter $|g\,\alpha_p|$, we have
%$[\tilde{f}'_k(\om), \tilde{f}^{\prime\dag}_k(\om)] \approx \gamma'_k\,\delta(\om-\om')$ with $\tilde{C}'_{kk}(\om)\approx1$.
%
If we were to redefine the noise operators as
$f_k(\om) \equiv (T_k)^{1/2}\,f'_k(\om)$ then
$C_{kk}(\om) \equiv T_k\,\tilde{C}_{kk}$ where
$\tilde{C}_{kk} = [f'_k(\om), f^{\prime\dag}_k(\om)] \approx \gamma'_k\,\delta(\om-\om')$ to lowest order in $|r_k|^2$.
This is the scaling employed by Raymer and McKinstrie \cite{Raymer:2013} for the intracavity fields
when comparing the operator equations of motions  in the high Q limit to the standard Langevin approach.

Other authors (see e.g., \cite{Tsang:2011, Shapiro:2015}) using the standard Langevin approach often simply state from the outset that $[f^{\prime\prime}_k(\om), f^{\prime\prime\dag}_{k}(\om')]=\delta(\om-\om')$, with the assumption that all cross commutators are zero (i.e. $C_{ab}=D_{ab}=0$), invoking independent noise sources.
One could, of course, obtain this form of the diagonal commutators by redefining
$f_k(\om) = (C_{kk})^{1/2}\,f^{\prime\prime}_k(\om)$, (with an appropriate rescaling of $H(\om)$).
However, even in the high Q limit we see from \Eq{solns:comms:Dab} that the cross commutator
$[f_a(\om), f_b(\om)]= D_{ab} \,\delta(\om-\om')$  remains non-zero (though small), unless we assume equal group velocities (index of refractions) for both the signal ($a$) and idler ($b$) modes so that $T_a = T_b$.

 \section{The output two-photon signal-idler state}\label{sec:2photon_state:outside}
 Inside the ring resonator the Hamiltonian in frequency space is
 \be{H_SPDC:om}
 \mathcal{H}^{NL} =  \int_{0_+}^{L_-}dz\int_{-\infty}^{\infty} \frac{d\om}{2\pi}\,
 g(\om)\,\left(\alpha_p(z,\om)\,a^\dag(z,\om)\,b^\dag(z,\om) + \alpha^*_p(z,\om)\,a(z,\om)\,b(z,\om) \right). %\quad 0_+\le z \le L_-.
 \ee
For a weak driving field $\alpha_p(\om) = |\alpha_p(\om)|\,\eithp$, the two-photon state inside the mrr is given by
 \bsub
 \bea{H_SPDC:state:ab}
 \ket{\Psi(T_{ab})}_{ab} &=& e^{-i\,\mathcal{H}^{NL}\,T_{ab}}\, \ket{\Psi}_{in} \approx \left(1 -i\,\mathcal{H}^{NL}\,T_{ab}\right)\ket{\vac} \\
 &=&
 \left[1 - i \int_{-\infty}^{\infty} \frac{d\om}{2\pi}\,r_{ab}(\om) \,\left( \eithp\,a^\dag(L_-,\om)\,b^\dag(L_-,\om) + \emithp\,a(L_-,\om)\,b(L_-,\om) \right)\right]\ket{\vac}, \qquad\quad \label{H_SPDC:state:ab:integral}
 \eea
 \esub
 where we have taken $T_{ab} = \sqrt{T_a T_b} = L/\sqrt{v_a\,v_b}$ assuming the group velocities of the generated signal and idler photons
 $v_a, v_b$ are not too different, and $|r_{ab}(\om)| \equiv |g(\om)\,\alpha_p(L_-,\om)\,T_{ab}|$.
 %
%% Original wording:
%The output state $\ket{\Psi}_{out}$ is obtained from $\ket{\Psi(T_{ab})}_{ab}$ by using the boundary condition equations \Eq{eqns:matrix:form:BD:out},
% which transforms $a(L_-,\om)\rightarrow a_{out}(\om)$ and $b(L_-,\om)\rightarrow b_{out}(\om)$
%and $r_{ab}(\om)\rightarrow g(\om)\,\alpha_{p,out}(\om)$ (which uses the value of $\langle c_{out}(\om)\rangle$ from \Eq{c:out})
%in \Eq{H_SPDC:state:ab:integral}.
 %
%% New wording: 24Mar2017
 For simplicity, in \Eq{H_SPDC:state:ab:integral} we have assumed perfect phase matching and zero dispersion.
 In general \cite{Scholz:2009, Schneeloch:2016, Schneeloch:2017, Sipe:2015a, Sipe:2015b}, when the field operators inside the mrr are decomposed in terms of their spatial Fourier components
 the spatial integral produces a phase matching contribution term, $\int_{0_+}^{L_-}dz \exp{[i\big(k_p(\om_p) - k_a(\om_a) - k_b(\om_b)\big)\,z}]$ for SPDC and
 $\int_{0_+}^{L_-}dz \exp{[i\big(2\,k_p(\om_p) - k_a(\om_a) - k_b(\om_b)\big)\,z}]$ for SFWM, yielding oscillatory $\sin$c function contributions over the longitudinal
 momentum conservation mismatch within the mrr. Further, dispersion effects within the mrr could be accounted for by Taylor expanding  $k(\om_k) = \om_k\,n_k(\om_k)/c$
 %using $v_k(\om_k) = c/n_k(\om)$
 about central frequencies $\om_{k,0}$ for $k\in\{p, a, b\}$ to either first or second order.
 While these spatially modulating $\sin$c factors (which are unity for perfect phase matching) and dispersion effects are important to account for in actual physical devices, we will ignore them here in this work for ease of exposition.

 The output state $\ket{\Psi}_{out}$ is obtained from the internal $\ket{\Psi(T_{ab})}_{ab}$ as the Heisenberg operators evolve from inside the mrr to the output bus.
 Making this substitution  $a(L_-,\om)\rightarrow a_{out}(\om)$ and $b(L_-,\om)\rightarrow b_{out}(\om)$ in \Eq{H_SPDC:state:ab:integral} and
 %
 %expression for $\ket{\Psi(T_{ab})}_{ab}$ in \Eq{H_SPDC:state:ab}.
 inserting the expressions for $a_{out}(\om)$ and $b_{out}(\om)$ from \Eq{aout:matrix:form} into \Eq{H_SPDC:state:ab:integral}  we obtain the output state
\bea{Psi:out:SPDC}
 \ket{\Psi}_{out} &=& \int_{-\infty}^{\infty} \frac{d\om}{2\pi}\,
 \left(
 \ket{\Psi^{(2)}(\om)}_{ab}\,\ket{\vac}_{env}
 \right. \no
 &-& i\,|\rab|\,
 \left[
  \ket{\phi_a^{(1)}(\om)}_a\,\ket{0}_b\,f_b^\dag(\om)\ket{\vac}_{env}
 + \ket{0}_a\,\ket{\varphi_b^{(1)}(\om)}_b\,f_a^\dag(\om)\,\ket{\vac}_{env}
 + \ket{0}_{ab}\,\ket{\Phi^{(2)}(\om)}_{env}
 \right], \qquad\;
\eea
where the vacuum state is given by $\ket{\vac} = \ket{0}_{ab}\ket{\vac}_{env} = \ket{0}_a\,\ket{0}_b\,\ket{\vac}_{env}$
such that $a_{in}\ket{0}_a = b_{in}\ket{0}_b = f_a\ket{0}_{env} = f_b\ket{0}_{env} = 0$.
The states in \Eq{Psi:out:SPDC} are given by
\bsub
\bea{Psi:out:terms}
\ket{\Psi^{(2)}(\om)}_{ab} &=& [2\,\pi\,\delta(\om) - i\,|\rab|\,C_{vac}(\om)]\,\ket{\vac}_{ab}
-i\,|\rab|\,\psi^{(2)}_{ab}(\om)\,a_{in}^\dag(\om)\,b_{in}^\dag(\om)\,\ket{0}_{ab}, \\
%
%&=& [2\,\pi\,\delta(\om) - i\,\rab\,C_{vac}(\om)]\,\ket{\vac}_{ab} -i\,r_p\,\psi^{(2)}_{ab}(\om)\,\ket{1_{\om},1_{-\om}}_{ab}, \\
%
C_{vac}(\om) &=&
\;\;\eithp\,
\big[
 G^*_{ab}(\om)\,G_{bb}(\om) + H^*_{ab}(\om)\,H_{bb}(\om)\,C_{bb}(\om)
\big] \no
&+&
\emithp\,
\big[
G_{aa}(\om)\,G^*_{ba}(\om) + H_{aa}(\om)\,H^*_{ba}(\om)\,C_{aa}(\om)
\big] \\
%H_{aa}(\om)\,H^*_{ba}(\om)\,C_{aa}(\om) + H^*_{ab}(\om)\,H_{bb}(\om)\,C_{bb}(\om) + G_{aa}(\om)\,G^*_{ba}(\om) + G^*_{ab}(\om)\,G_{bb}(\om),
%
\psi^{(2)}_{ab}(\om) &=& \eithp\,G^*_{aa}(\om)\,G_{bb}(\om) + \emithp\,G_{ab}(\om)\,G^*_{ba}(\om), \label{psi2}\\
\ket{\phi_a^{(1)}(\om)}_a &=& \left[\eithp\,G^*_{aa}(\om)\,H_{bb}(\om) + \emithp\,G^*_{ba}(\om)\,H_{ab}(\om)\right] \, a_{in}^\dag(\om)\,\ket{0}_a, \no
&\equiv&  \phi_a^{(1)}(\om)\,\ket{1_{\om}}_a, \\
\ket{\varphi_b^{(1)}(\om)}_b &=& \left[\eithp\,G_{bb}(\om)\,H^*_{aa}(\om) + \emithp\,G_{ab}(\om)\,H^*_{ba}(\om)\right]\, b_{in}^\dag(\om)\,\ket{0}_b, \no
&\equiv&  \varphi_b^{(1)}(\om)\,\ket{1_{-\om}}_b, \\
\ket{\Phi^{(2)}(\om)}_{env} &=& \left[\eithp\, H^*_{aa}(\om)\,H_{bb}(\om)\,f^\dag_a(\om)\,f^\dag_b(\om) + \emithp\, H_{ab}(\om)\,H^*_{ba}(\om)\,f^\dag_b(\om)\,f^\dag_a(\om)\right]\,\ket{\vac}_{env}.\qquad\;
\eea
\esub
In the above,  $C_{vac}(\om)$ is the first order (in $|\rab|$) correction to the signal-idler vacuum state, and $|r_{ab}(\om)|\,\psi^{(2)}_{ab}(\om)$ is the two photon wavefunction.
 From \Eq{psi2} and \Eq{G:HighQ:limit:1storder} we observe that to zeroth order in $|g\,\alpha_p\,T_{ab}|$, the output two-photon state
 $\psi^{(2)}_{ab}(\om)\approx \eithp\,G^*_{aa}(\om)\,G_{bb}(\om)$ involves the frequency dependent shifts of the input fields to the output fields.
 The second term in \Eq{psi2} $\emithp\,G_{ab}(\om)\,G^*_{ba}(\om)\propto |g\,\alpha_p\,T_{ab}|^2$
 represents a higher order pump dependent correction to $\psi^{(2)}_{ab}(\om)$
 involving the product of Lorentzian lineshape factors
$\sqrt{\gamma_k}/(s+\G_k/2)$
(where $s\equiv -i\om$ can be considered as a Laplace transform solution variable \cite{Tsang:2011}),
relating the fields $(\vec{a})_k$
inside the cavity to the input fields $(\vec{a}_{in})_k$.

We are interested in the reduced density matrix of the signal-idler system obtained from
$\rho_{ab}~=~\Tr_{env}[\ket{\Psi}_{out}\bra{\Psi}]$. To trace over the environment we use the fact that
$\Tr_{env}\left[f^\dag_{i'}(\om')\ket{\vac}_{env}\bra{\vac}\,f_{i}(\om)\right]$
$={}_{env}\bra{\vac}\,f_{i}(\om)\,f^\dag_{i'}(\om')\,\ket{\vac}_{env} = C_{i\,i'}\,\delta(\om -\om')$,
where we have used $f_{i'}\,f^\dag_i = [f_{i'}, f^\dag_i] +f^\dag_i\, f_{i'}$.
Similarly,
$\Tr_{env}\left[f^\dag_{i'}(\om')\,f^\dag_{j'}(\om')\,\ket{\vac}_{env}\bra{\vac}\,f_{i}(\om)\,f_{j}(\om)\right]=$
${}_{env}\bra{\vac}\,f_{i}(\om)\,f_{j}(\om)\,f^\dag_{i'}(\om')\,f^\dag_{j'}(\om') \,\ket{\vac}_{env} = [C_{i\,i'}(\om)\,C_{j\,j'}(\om) + C_{i\,j'}(\om)\,C_{j\,i'}(\om)]\,\delta(\om-\om')$.
Using the additional fact that $C_{ab}(\om)=0$ from \Eq{comms:spdc:abdag}, we have
\bsub
\bea{rho:ab:spdc}
\rho_{ab} &=& \int_{-\infty}^{\infty} \frac{d\om}{2\pi}\,\ket{\Psi^{(2)}(\om)}_{ab} \bra{\Psi^{(2)}(\om)}
+ |\rab|^2 \, \left( \int_{-\infty}^{\infty} \frac{d\om}{2\pi}\,R_0(\om) \ket{0}_{ab}\bra{0}, \right.  \\
&+&  \left. \int_{-\infty}^{\infty} \frac{d\om}{2\pi}\,|\phi_a^{(1)}(\om)|^2\,C_{bb}(\om)\,\ket{1_{\om},0}_{ab}\bra{1_{\om},0}
 +    \int_{-\infty}^{\infty} \frac{d\om}{2\pi}\,C_{aa}(\om)\,|\varphi_b^{(1)}(\om)|^2\,\ket{0,1_{\om}}_{ab}\bra{0,1_{\om}} \right),  \no
%
%&+& \int_{-\infty}^{\infty} \frac{d\om}{2\pi}\,R_0(\om) \ket{0}_{ab}\bra{0}, \\
%
\ket{\Psi^{(2)}(\om)}_{ab} &=& [2\,\pi\,\delta(\om) - i\,|\rab|\,C_{vac}(\om)]\,\ket{0}_{ab} -i\,|\rab|\,\psi^{(2)}_{ab}(\om)\,\ket{1_{\om},1_{-\om}}_{ab}, \\
R_0(\om) &=& C_{ab}(\om)\,C_{ab}(\om)\,|\eithp\,H^*_{aa}(\om)\,H_{bb}(\om) + \emithp\,H_{ab}(\om)\,H^*_{ba}(\om))|^2.
\eea
\esub
In the above, $\ket{\Psi^{(2)}(\om)}_{ab}$ is the two-photon signal-idler state including the vacuum.
The two-photon generation rate \cite{Scholz:2009,Tsang:2011} is given by
$R_{ab}(\om) = |\rab|^2\,|\psi^{(2)}_{ab}(\om)|^2$. The second line of \Eq{rho:ab:spdc} gives the single photon contributions due to loss of a idler (leftmost term) or signal photon (rightmost term) with singles rates $|\rab|^2\,|\phi_a^{(1)}(\om)|^2\,C_{bb}(\om)$ and $|\rab|^2\,C_{aa}(\om)\,|\varphi_b^{(1)}(\om)|^2$ respectively, where effect of the noise commutators are explicitly evident.
We can therefore write
\be{rho:b}
\rho_{ab} =  \textrm{Tr}_{env}\left[ \ket{\Psi}_{out}\bra{\Psi}\right] = \sum_{k=0,1,2} p_k\,\rho_{ab}^{(k)}, \quad
\textrm{Tr}_{ab}[ \rho_{ab}^{(k)} ]=1, \quad \sum_{k=0,1,2} p_k=1.
\ee
Here $\rho_{ab}^{(k)}$ with $k\in\{0, 1, 2\}$ represents the $k$ \textit{system}-photon (i.e. signal-idler)
portion of the reduced density matrix $\rho_{ab}$.
One can then define the output coincidence to accidental rate (CAR) \cite{Tsang:2011} as
\bea{RCAR:out}
R^{(out)}_{\textrm{CAR}}(\om) &=&  \frac{|\psi^{(2)}_{ab}(\om)|^2}{|\phi_a^{(1)}(\om)|^2\,C_{bb}(\om) + C_{aa}(\om)\,|\varphi_b^{(1)}(\om)|^2} \no
&\myover{ {\longrightarrow} {\stackrel{\textrm{\scriptsize high Q}}{\mathcal{O}(|g\,\alpha_p|)}}}&
\fracd{|G^*_{aa}(\om)\,G_{bb}(\om)|^2}
{|G^*_{aa}(\om)\,\tilde{H}_{bb}(\om)|^2+|\tilde{H}^*_{aa}(\om)\,G_{bb}(\om)|^2} =
\left( \fracd{\g_a\,\g'_a}{\om^2 + (\Delta_a/2)^2} + \fracd{\g_b\,\g'_b}{\om^2 + (\Delta_b/2)^2}\right)^{-1}, \qquad\quad
\eea
and the output heralding efficiency \cite{Tsang:2011} of say the an idler photon by the measurement of a signal photon as
\bea{Rherald:out}
R^{(out)}_{\textrm{herald}}(\om) &=& \frac{|\psi^{(2)}_{ab}(\om)|^2}{|\phi_a^{(1)}(\om)|^2\,C_{bb}(\om) + |\psi^{(2)}_{ab}(\om)|^2} \no
&\myover{ {\longrightarrow} {\stackrel{\textrm{\scriptsize high Q}}{\mathcal{O}(|g\,\alpha_p|)}}}&
\fracd{|G^*_{aa}(\om)\,G_{bb}(\om)|^2}
{|G^*_{aa}(\om)\,\tilde{H}_{bb}(\om)|^2+|G^*_{aa}(\om)\,G_{bb}(\om)|^2} =
\left( 1 + \fracd{\g_b\,\g'_b}{\om^2 + (\Delta_b/2)^2}\right)^{-1},
\eea
where we have used
$H_{kk}(\om)\,C_{kk}\rightarrow H_{kk}(\om)\,(1-\alpha_k^2)^{1/2} \rightarrow \g'_k\,H_{kk}(\om) = \tilde{H}_{kk}(\om)$ in the high cavity Q limit.
Note that in the first lines in \Eq{RCAR:out} and \Eq{Rherald:out} a common factor of $|\rab|^2 = |g\,\alpha_p\,T_{ab}|^2$ in the numerator and denominator has been canceled.

\section{The two-photon signal-idler state inside the mrr}\label{sec:2photon_state:inside_mrr}
It is noteworthy to investigate the state of the two-photon state \textit{inside} the mrr cavity, since it is this state which is most often computed in other treatments \cite{Scholz:2009,Tsang:2011}
(with the output field usually given as simply
%$(\gamma_a\,\gamma_b)^{1/2}$
$\sqrt{\gamma_a\,\gamma_b}$
times the input field (see e.g., \cite{Walls_Milburn:1994,Scholz:2009}).

For a weak driving field $\alpha_p(\om) = |\alpha_p(\om)|\,\eithp$ the two-photon state inside the mrr is given by \Eq{H_SPDC:state:ab}, which for convenience we restate below,
 \bsub
 \bea{H_SPDC:state:abL}
 \ket{\Psi(T_{ab})}_{ab} &=& e^{-i\,\mathcal{H}^{NL}\,T_{ab}}\, \ket{\Psi}_{in} \approx \left(1 -i\,\mathcal{H}^{NL}\,T_{ab}\right)\ket{\vac} \no
 &=&
 \left[1 - i \int_{-\infty}^{\infty} \frac{d\om}{2\pi}\,r_{ab}(\om) \,\left( \eithp\,a^\dag(L_-,\om)\,b^\dag(L_-,\om) + \emithp\,a(L_-,\om)\,b(L_-,\om) \right)\right]\ket{\vac}. \qquad\quad \label{H_SPDC:state:abL:integral} \nonumber
 \eea
 \esub

 Using the output boundary condition \Eq{eqns:matrix:form:BD:out}, and \Eq{G:H} relating the output fields to the input fields one obtains
 \be{aL}
 \vec{a}(L_-,\om) = \left[X^{-1}_\tau\,H(\om)\,P_{\chi}\,X_\tau\right]\,\vec{a}_{in} + \left[ X^{-1}_\tau\,H(\om)\right]\,\vec{f}(\om)
 \equiv G^{(L)}\,(\om)\,\vec{a}_{in}(\om) + H^{(L)}\,(\om)\,\vec{f}(\om),
 \ee
 with (employing the expression for $H(\om)$ in \Eq{Hom:v2})
%===========================
% G(L) p9.3 22Mar2017 notes
%===========================
\bsub
\bea{GLom}
G^{(L)}(\om)
&=& \frac{1}{D(s)} \,
\left(
  \begin{array}{cc}
    \tau_a\,(1-\rho_b\exib)\,\exia  & -i\,r_a\,\tau_b\,\exib \\
     i\,r_b\,\tau_a\,\exia          & \tau_b\,(1-\rho_a\exia)\,\exib
  \end{array}
\right), \\
&\myover{ {\longrightarrow} {\stackrel{\textrm{\scriptsize high Q}}{\mathcal{O}(|g\,\alpha_p|)}}} &
\left(
  \begin{array}{cc}
   \fracd{\sqrt{\gamma_a}}{s+\G_a/2}   & -i\,r_a\, \left(\fracd{1}{s+\G_a/2}\right)\, \left(\fracd{\sqrt{\gamma_b}}{s+\G_b/2}\right)\,\exib \\
   -i\,r_b\,\left(\fracd{\sqrt{\gamma_a}}{s+\G_a/2}\right)\, \left(\fracd{1}{s+\G_b/2}\right)\,\exia  & \fracd{\sqrt{\gamma_b}}{s+\G_b/2},
  \end{array}
\right), \label{GLom:highQ:firstorder}\qquad\quad
\eea
\esub
where in \Eq{GLom:highQ:firstorder} we have used $\exik \approx 1$,
and
%==========
% H better
%==========
\bsub
\bea{HLom}
H^{(L)}(\om)
\hspace*{-1em}
&=&
\hspace*{-1em}
\frac{1}{D(s)} \,
\left(
  \begin{array}{cc}
    (1-\rho_b\exib)  & -i\,r_a \\
     i\,r_b\         & (1-\rho_a\exia)
  \end{array}
\right) \equiv \tilde{H}^{(L)}(\om)\,\Lambda^{-1}_{\alpha}(\om), \;
\Lambda_{\alpha} \equiv
\left(
  \begin{array}{cc}
   (1-\alpha_a^2)^{1/2}  & 0 \\
    0 & (1-\alpha_b^2)^{1/2} \\
  \end{array}
\right), \quad\quad\; \\
&\myover{ {\longrightarrow} {\stackrel{\textrm{\scriptsize high Q}}{\mathcal{O}(|g\,\alpha_p|)}} }& %&{}&
\left(
  \begin{array}{cc}
   \fracd{1}{s+\G_a/2}   &-i\,r_a\, \left(\fracd{1}{s+\G_a/2}\right)\, \left(\fracd{1}{s+\G_b/2}\right)\,\exib \\
   -i\,r_b\,\left(\fracd{1}{s+\G_a/2}\right)\, \left(\fracd{1}{s+\G_b/2}\right)\,\exia & \fracd{1}{s+\G_b/2},
  \end{array} \label{HLom:highQ:firstorder}
\right) \qquad\qquad
\eea
\esub
where we have similarly defined $\tilde{H}^{(L)}(\om)$   as in \Eq{aout:tilde:H}.
The calculation of the wavefunction $\ket{\Psi(T_{ab})}_{ab}$ and reduced density matrix $\rho_{ab}(T_{ab})$ inside the mrr at $z=L_-$
proceeds identically as in \Sec{sec:2photon_state:inside_mrr} except for the replacement of $G(\om)\rightarrow G^{(L)}(\om)$ and
$H(\om)\rightarrow H ^{(L)}(\om)$ in \Eq{Psi:out:SPDC} and \Eq{rho:ab:spdc} respectively.
Analogous to \Eq{psi2}, the two-photon wavefunction inside the mrr is given by $|r_{ab}(\om)|$ times
\be{biphoton:wavefunction}
\psi^{(2)}_{ab,\, mrr}(\om) = \eithp\,G^{(L)*}_{aa}(\om)\,G^{(L)}_{bb}(\om) + \emithp\,G^{(L)}_{ab}(\om)\,G^{(L)*}_{ba}(\om).
\ee
To zeroth order in $|g\,\alpha_p\,T_{ab}|$, the first term gives
$\psi^{(2)}_{ab,\, mrr}(\om)\approx\,G^{(L)*}_{aa}(\om)\,G^{(L)}_{bb}(\om) = \left[\sqrt{\gamma_a}/(s+\G_a/2)\right]\,$ $\left[\sqrt{\gamma_b}/(s+\G_b/2)\right]$,
the product of the standard Langevin input-output theory Loretzian lineshape factors for each field $a, b$.
This is the typical expression found in other works computing the two photon state
inside a cavity or mrr \cite{Scholz:2009, Tsang:2011, Chen:2011, Camacho:2012}.
The starting point for many such calculations invoking the standard Langevin input-output  formalism \cite{Walls_Milburn:1994, Orszag:2000}
begins with the statement
that the (generic) free field operator $a(\om)$ is modified inside the cavity or mrr by the change in the density of states, which is accounted for by the substitution
$a(\om) \rightarrow \sqrt{\gamma_a}\,a(\om)/(s+\G_a/2)$.
Again, the second term $\emithp\,G^{(L)}_{ab}(\om)\,G^{(L)*}_{ba}(\om)$ in $\psi^{(2)}_{ab,\, mrr}(\om)$ represents a second order (in $|g\,\alpha_p|$) pump-dependent correction.
Inside the mrr cavity, the expressions for the coincidence to accidental rate (CAR)
\bea{RCAR:mrr}
R^{(mrr)}_{\textrm{CAR}}(\om) &=&  \frac{|\psi^{(2)}_{ab,\, mrr}(\om)|^2}{|\phi_{a,\, mrr}^{(1)}(\om)|^2\,C_{bb}(\om) + C_{aa}(\om)\,|\varphi_{b,\, mrr}^{(1)}(\om)|^2} \no
&\myover{ {\longrightarrow} {\stackrel{\textrm{\scriptsize high Q}}{\mathcal{O}(|g\,\alpha_p|)}}}&
\fracd{|G^{(L)*}_{aa}(\om)\,G^{(L)}_{bb}(\om)|^2}
{|G^{(L)*}_{aa}(\om)\,\tilde{H}^{(L)}_{bb}(\om)|^2+|\tilde{H}^{(L)*}_{aa}(\om)\,G^{(L)}_{bb}(\om)|^2}=
\frac{\g_a\,\g_b}{\g_a\,\g'_a + \g_b\,\g'_b},
\eea
and the  heralding efficiency of say an idler photon by the measurement of a signal photon as
\bea{Rherald:mrr}
R^{(mrr)}_{\textrm{herald}}(\om) &=& \frac{|\psi^{(2)}_{ab,\, mrr}(\om)|^2}{|\phi_{a,\, mrr}^{(1)}(\om)|^2\,C_{bb}(\om) + |\psi^{(2)}_{ab,\, mrr}(\om)|^2} \no
&\myover{ {\longrightarrow} {\stackrel{\textrm{\scriptsize high Q}}{\mathcal{O}(|g\,\alpha_p|)}}}&
\fracd{|G^{(L)*}_{aa}(\om)\,G^{(L)}_{bb}(\om)|^2}
{|G^{(L)*}_{aa}(\om)\,\tilde{H}^{(L)}_{bb}(\om)|^2+|G^{(L)*}_{aa}(\om)\,G^{(L)}_{bb}(\om)|^2}=
\frac{\g_a\,\g_b}{\g_a\,\g'_b + \g_a\,\g_b} = \frac{\g_b}{\G_b}.
\eea
\Eq{RCAR:mrr} and \Eq{Rherald:mrr} generalize the expressions of Tsang \cite{Tsang:2011} which were computed for a cavity using the standard Langevin input-output formalism
(recalling that to lowest order in
$|g\,\alpha_p\,T_{ab}|$ we have $C_{kk}\approx \gamma'_k$ for $k\in\{a, b\}$ so that $H^{(L)}_{bb}(\om)\,C_{kk}\rightarrow\tilde{H}^{(L)}_{kk}(\om)$).
Both of the above expressions
%\Eq{RCAR:mrr} and \Eq{Rherald:mrr}
suggest that the minimization of internal propagation losses  $\g'_k\ll\g_k$,
is desirable for the generation of pure entangled photons.

\begin{figure}[ht]
\begin{tabular}{cc}
\includegraphics[width=3.25in,height=2.25in]{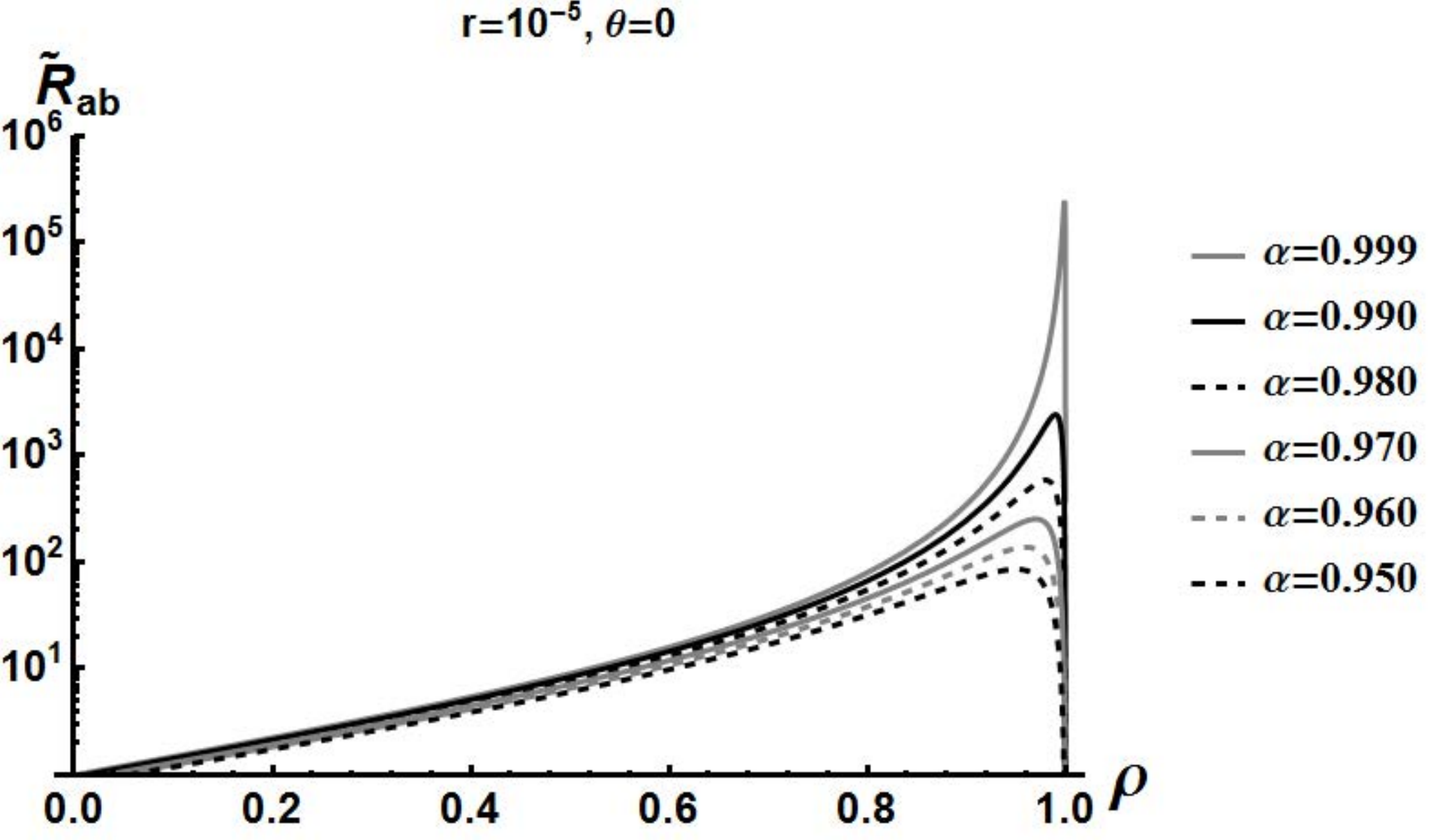} & %{fig2_ppg_left_Rtilde_r1e-5_theta0_alphas_v3}
\includegraphics[width=3.25in,height=2.25in]{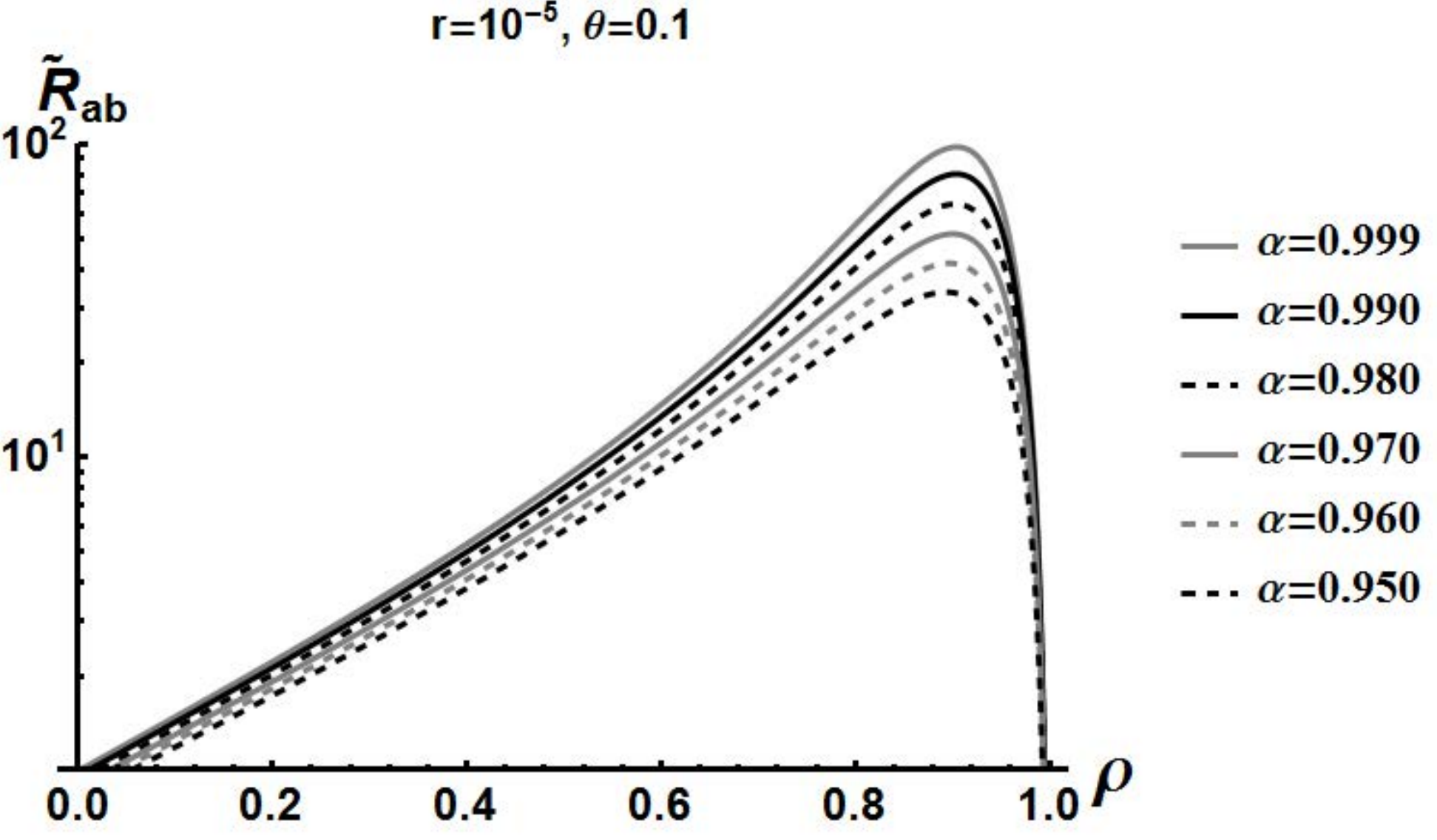}%{fig2_ppg_right_Rtilde_r1e-5_theta0p1_alphas_v3}
\end{tabular}
\caption{$\tilde{R}^{(mrr)}_{ab}=|\psi^{(2)}_{ab,mrr}(\omega)|^2 =R^{(mrr)}_{ab}/|\rab|^2$ for
$r=10^{-5}$ and (left) on mrr resonance $\theta = 0$, (right) slightly off mrr resonance $\theta = 0$, for $\alpha = (0.999, 0.99, 0.98, 0.97, 0.95, 0.95)$.
}\label{fig:Rtilde:r1e-5:theta0:alphas}
\end{figure}

\begin{figure}[ht]
%\begin{tabular}{cc}
\includegraphics[width=3.5in,height=2.5in]{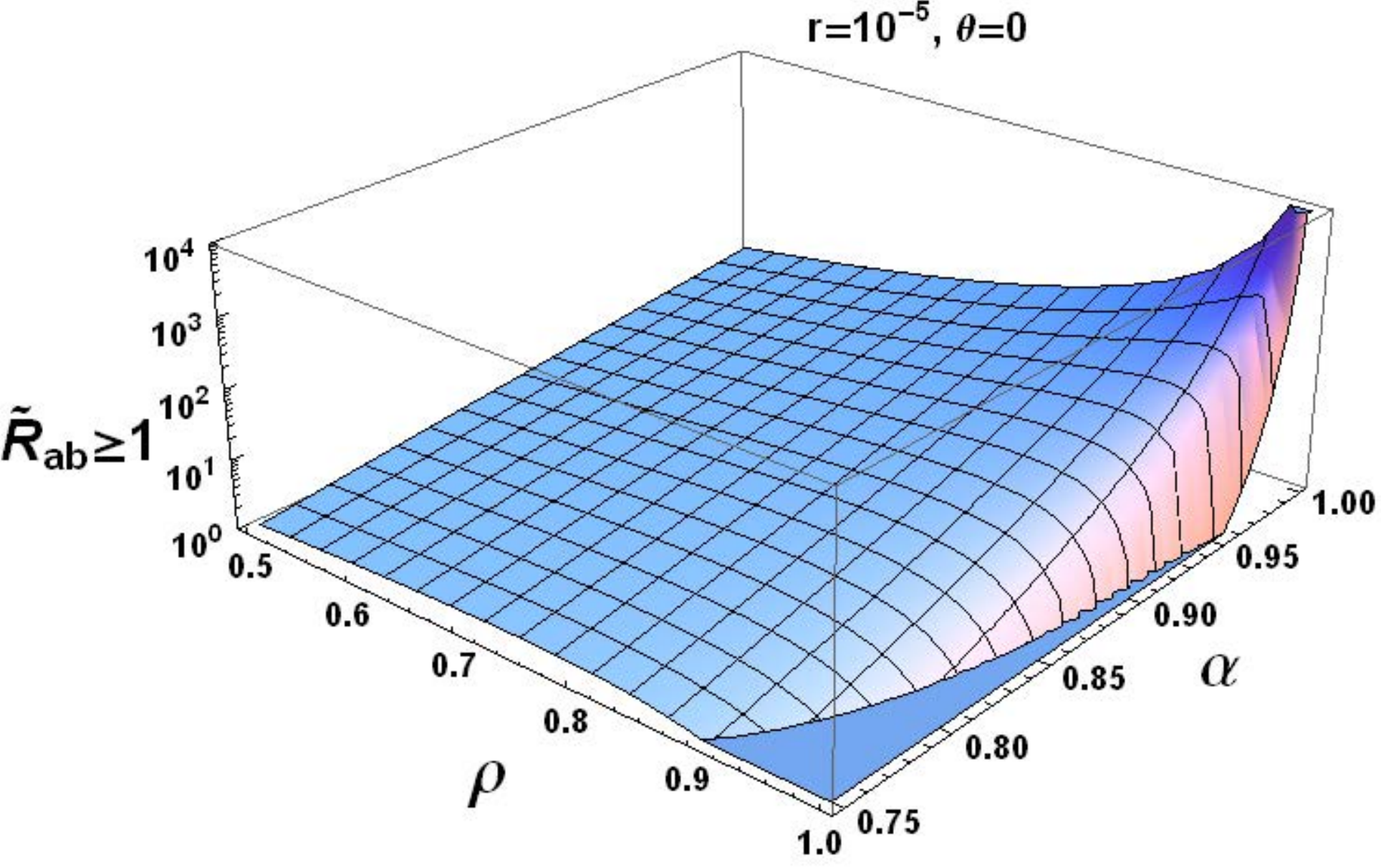}%{fig3_ppg_Rtilde3D_r1e-5_theta0_rhos_alphas_v3} %&
%\includegraphics[width=3.0in,height=2.25in]{Rtilde_r1e-5_rho0p5_theta_alphas.png}
%\end{tabular}
\caption{$\tilde{R}^{(mrr)}_{ab}=|\psi^{(2)}_{ab,mrr}(\omega)|^2$ for
$r=10^{-5}$ on mrr resonance $\theta=0$ for $0.5\le\rho\le1.0$ and $0.75\le\alpha\le1.0$.
}\label{Rtilde3D:r1e-5:theta0:rhos:alphas}
\end{figure}

\begin{figure}[ht]
\begin{tabular}{cc}
\includegraphics[width=3.25in,height=2.25in]{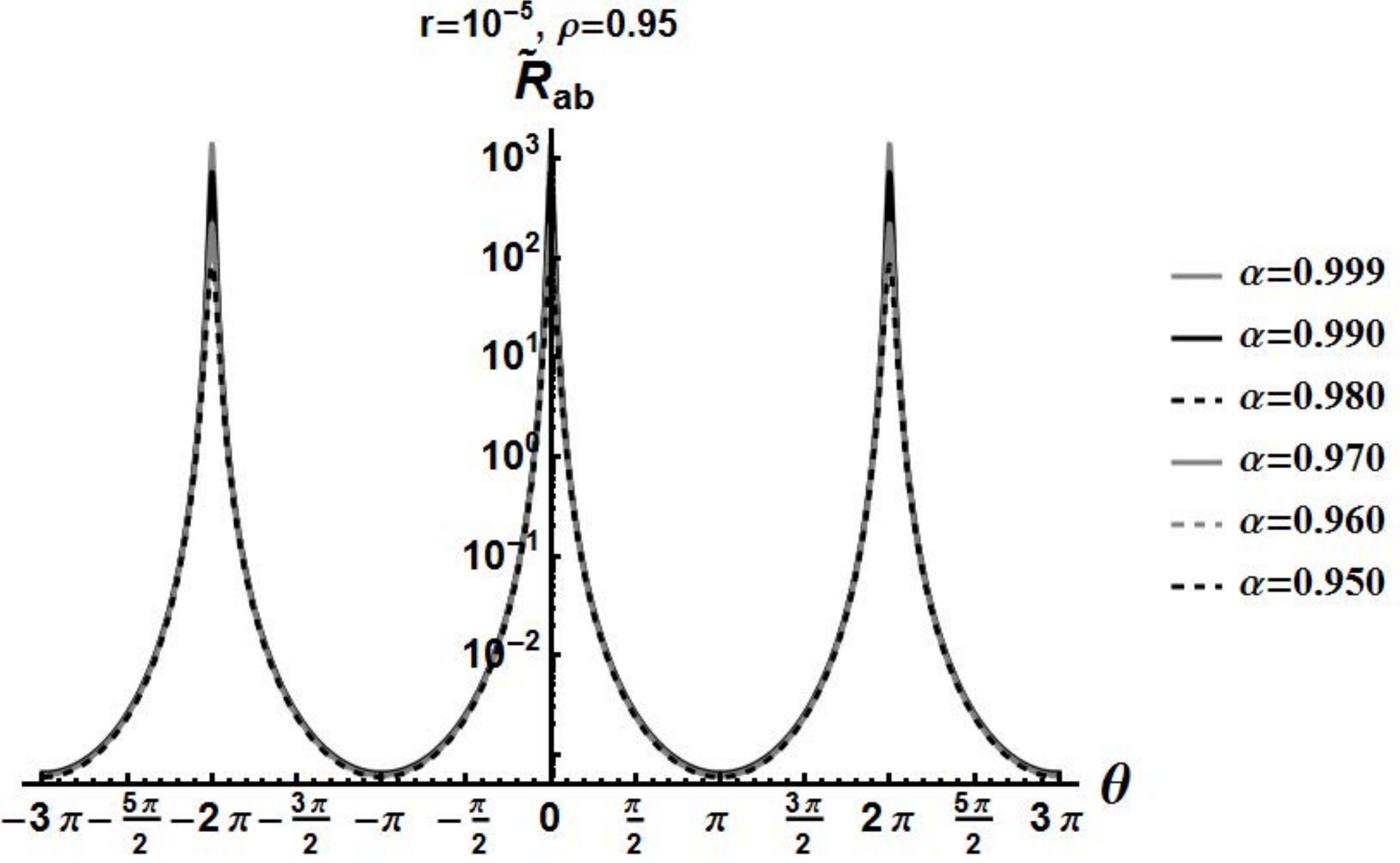} & %{fig4_ppg_left_Rtilde_r1e-5_rho0p95_theta_alphas_v3}
\includegraphics[width=3.25in,height=2.25in]{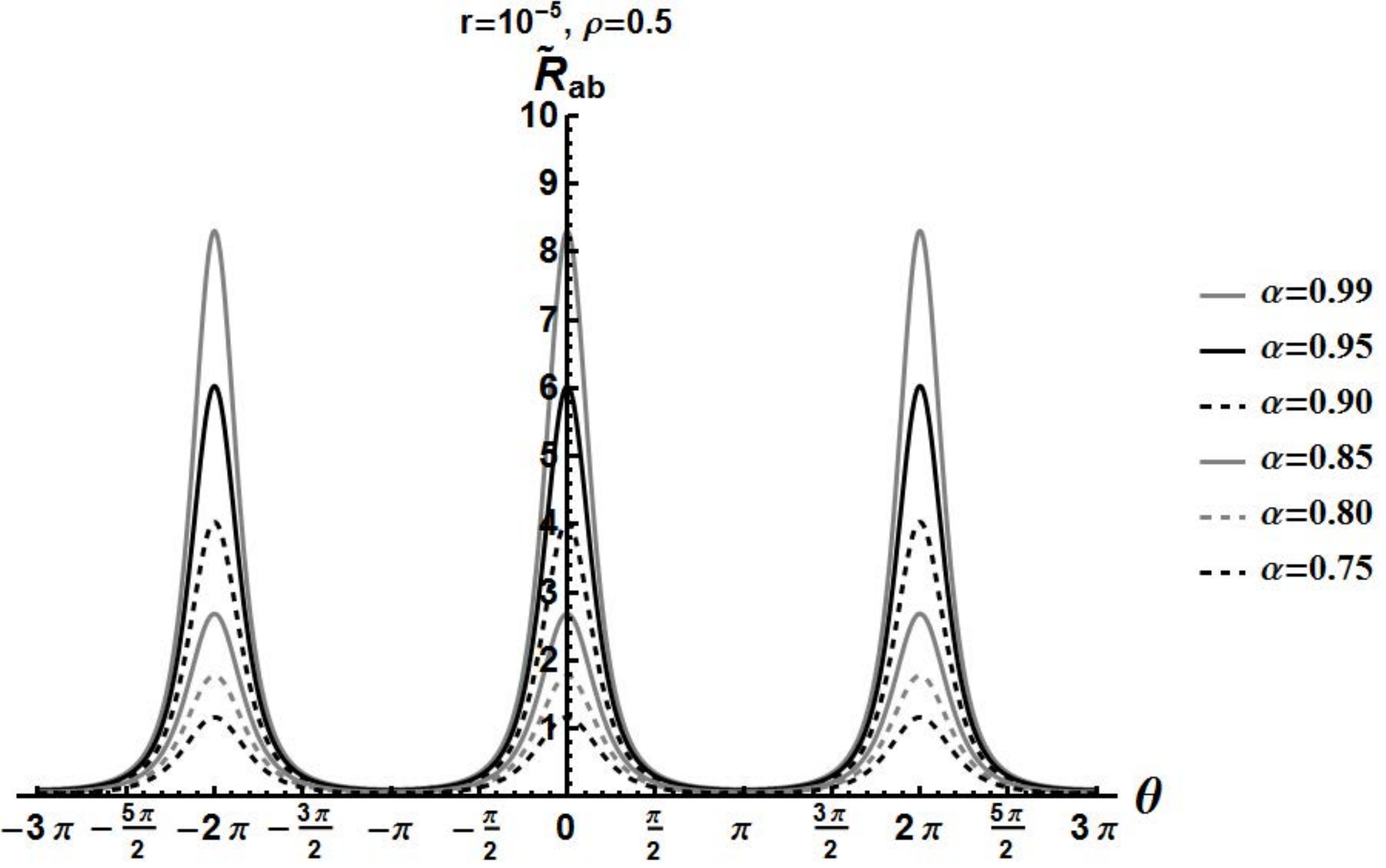}%{fig4_ppg_right_Rtilde_r1e-5_rho0p5_theta_alphas_v3}
\end{tabular}
\caption{$\tilde{R}^{(mrr)}_{ab}=|\psi^{(2)}_{ab,mrr}(\omega)|^2$ for
$r=10^{-5}$ and (left) $\rho = 0.95$, and $\alpha = (0.999, 0.99, 0.98, 0.97, 0.95, 0.95)$, and
(right) (right) $\rho = 0.5$ for $\alpha = (0.99, 0.95, 0.90, 0.85, 0.80, 0.75)$
}\label{Rtilde:r1e-5:rho0p95:theta:alphas}
\end{figure}

The expression for the biphoton production rate inside the mrr is given by
\be{biphoton:prod:rate}
R^{(mrr)}_{ab} = |\rab|^2 \,|\psi^{(2)}_{ab,mrr}(\omega)|^2
\ee
where the two-photon wavefunction inside the mrr is given by
\bea{psi2abmrr}
\psi^{(2)}_{ab,\, mrr}(\om) &=&
\eithp\,G^{(L)*}_{aa}(\om)\,G^{(L)}_{bb}(\om) + \emithp\,G^{(L)}_{ab}(\om)\,G^{(L)*}_{ba}(\om), \no
&=&
\frac{ \alpha_a \alpha_b \, \tau_a \tau_b \,e^{i \left(\theta_b+\theta _p\right)}\,
\left[e^{i \theta _b} |r_a|\,|r_b|
-\left(1-e^{i \theta _a} \alpha _a \rho _a\right)
\left(e^{i \theta _b}-\alpha _b \rho _b\right)
\right] }
% denom
{\left[
  e^{i \left(\theta_a+\theta _b\right)}
  |r_a|\,|r_b|-(e^{i \theta _a}-\alpha_a \rho _a)\, (e^{i \theta_b}-\alpha_b \rho_b)
\right]\,
\big[
  (1-e^{i \theta_a} \alpha _a \rho_a)\,(1-e^{i\theta_b} \alpha_b \rho_b)-|r_a|\,|r_b|
\big]}. \qquad
\eea
In \Fig{fig:Rtilde:r1e-5:theta0:alphas} we plot
$\tilde{R}^{(mrr)}_{ab}=|\psi^{(2)}_{ab,mrr}(\omega)|^2 =R^{(mrr)}_{ab}/|\rab|^2$ for a
weak driving pump $|r_a|=|r_b|=r=10^{-5}$ on mrr resonance $\theta=\om\,T=0$, and slightly off resonance at $\theta=0.1$.
In this (and subsequent) plot(s), we have considered equal mrr round trip  times $T_a=T_b=T$ for both the signal and idler so that $\theta_a=\theta_b\equiv\theta=\om\,T$, as well as equal coupling $\rho_a=\rho_b\equiv\rho$, and internal loss $\alpha_a=\alpha_b\equiv\alpha$.
Here $\alpha=(0.99, 0.95)$ represents the physically relevant values of $1\%$ and $5\%$ propagation loss within the mrr, respectively.
%=======================================================
%=======================================================
Note that $\tilde{R}^{(mrr)}_{ab}$ is independent of the pump phase
$\theta_p$ as can be observed from the overall factor of $\eithp$ in \Eq{psi2abmrr}.
The surface of $\tilde{R}^{(mrr)}_{ab}\ge1$ for resonance $\theta=0$ as a function of coupling $\rho$ and  internal propagation loss $\alpha$ is plotted in \Fig{Rtilde3D:r1e-5:theta0:rhos:alphas}.
This plot indicates that strong biphoton pair production is favored by a high cavity Q ($\rho\rightarrow 1$), and low internal propagation loss ($\alpha\rightarrow 1$).
%=======================================================
%=======================================================
%
In \Fig{Rtilde:r1e-5:rho0p95:theta:alphas} we  plot $\tilde{R}^{(mrr)}_{ab}$ as a function of $\theta=\om\,T$
for $\rho=0.95$ and $\rho=0.50$, where the effect of the resonance structure of the mrr is manifest.
%=======================================================
%=======================================================
In the \App{app}
%\App{app:highQ:limit:Rab}
we compare the expression for the biphoton generation rate in the high cavity Q limit with
other expressions derived in the literature \cite{Scholz:2009,Tsang:2011} using the standard Langevin approach.

%=======================================================

The expressions for $R^{(mrr)}_{\textrm{CAR}}(\om)$ in \Eq{RCAR:mrr} and $R^{(mrr)}_{\textrm{herald}}(\om)$ in \Eq{Rherald:mrr} take on simple analytic forms given by
\be{RCAR:mrr:expression}
R^{(mrr)}_{\textrm{CAR}}(\om) =
\fracd{\alpha_a^2\,\alpha_b^2\,\tau_a^2\,\tau_a^2\,}
{(1-|r_a|^2)\,\alpha_b^2\,\alpha_b^2 +\alpha_a^2\,\left[(1-|r_b|^2 - \alpha_b^2)\tau_a^2 -\alpha_b^2\,\alpha_b^2\right]}
\rightarrow
\fracd{\alpha^2\,(1-\rho^2)}{2\,(1-|r|^2 - \alpha^2)},
\ee
were in the last expression we have again used $T_a=T_b=T$, $\rho_a=\rho_b=\rho$ and $\alpha_a=\alpha_b=\alpha$.
Note $R^{(mrr)}_{\textrm{CAR}}(\om)$ is independent of $\theta_a,\,\theta_b$.
%=======================================================
\begin{figure}[ht]
%\begin{tabular}{cc}
\includegraphics[width=4.5in,height=2.5in]{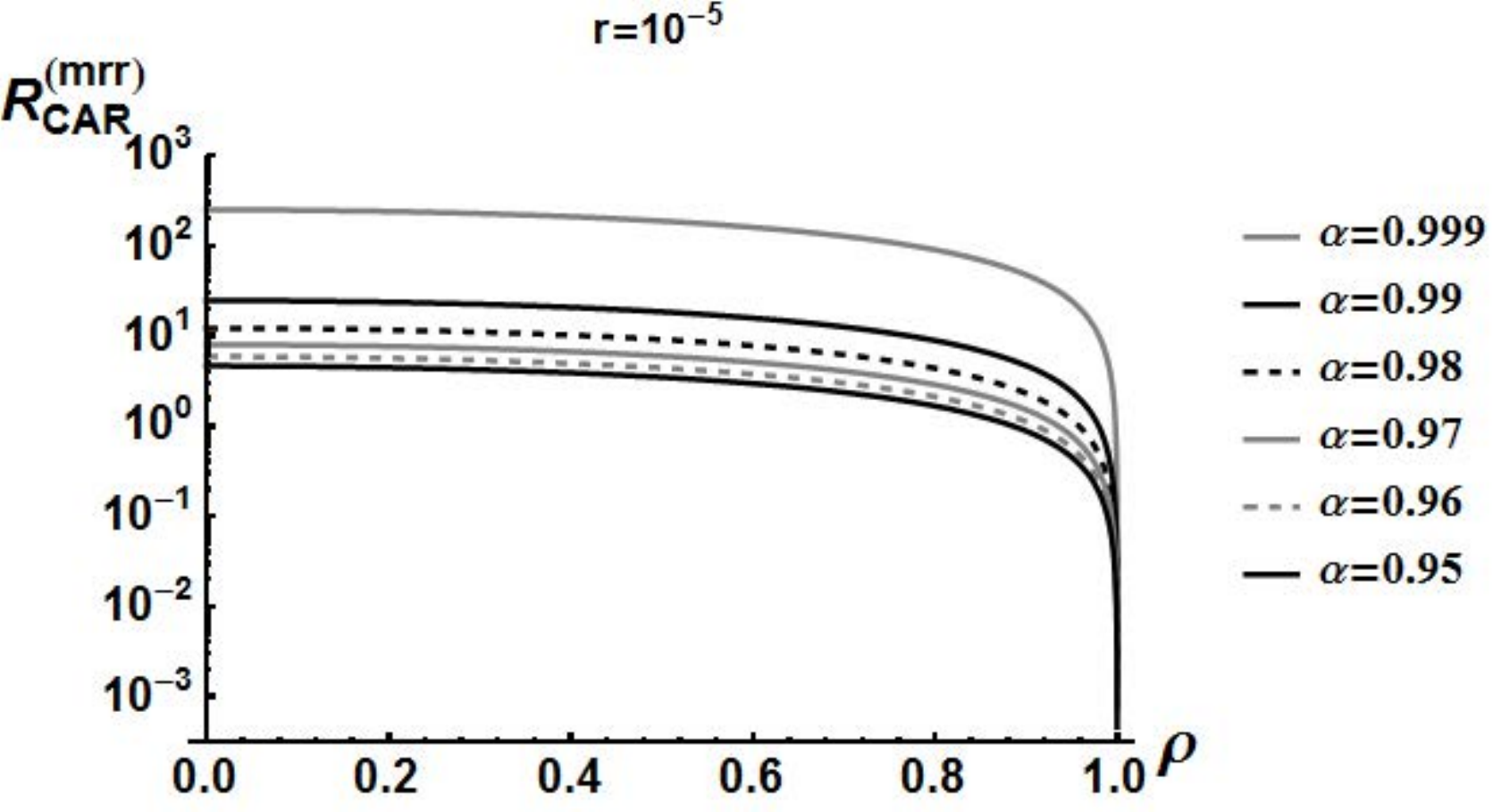}%{fig5_ppg_RCARmrr_r1e-5_rho_alphas_27Jul2017_v2.png} %&
%\includegraphics[width=4.5in,height=2.5in]{fig5_ppg_RCARmrr_r1e-5_rho_alphas_v3} %&
%\includegraphics[width=3.0in,height=2.25in]{Rtilde_r1e-5_rho0p5_theta_alphas.png}
%\end{tabular}
\caption{Coincidence to accidental rate $R^{(mrr)}_{\textrm{CAR}}(\om)$ with
$r=10^{-5}$ for $\alpha = (0.999, 0.99, 0.98, 0.97, 0.96, 0.95)$.
}\label{RCARmrr:r1e-5:rho:alphas}
\end{figure}
%=======================================================
In \Fig{RCARmrr:r1e-5:rho:alphas} we plot $R^{(mrr)}_{\textrm{CAR}}(\om)$ with
$r=10^{-5}$ for the operationally relevant (for $\alpha\le 0.99$) internal propagation loss values $\alpha = (0.999, 0.99, 0.98, 0.97, 0.95, 0.95)$.

The heralding efficiency $R^{(mrr)}_{\textrm{herald}}(\om)$ takes even a simpler form, which again is independent of the phase accumulation angle $\theta_b$
\be{Rherald:expression}
R^{(mrr)}_{\textrm{herald}}(\om) =
\fracd{\alpha_b^2\,(1-\rho_b^2)}{(1-|r_b|^2 - \alpha_b^2\,\rho_b^2)}.
\ee
%=======================================================
\begin{figure}[ht]
\begin{tabular}{cc}
\includegraphics[width=3.5in,height=2.5in]{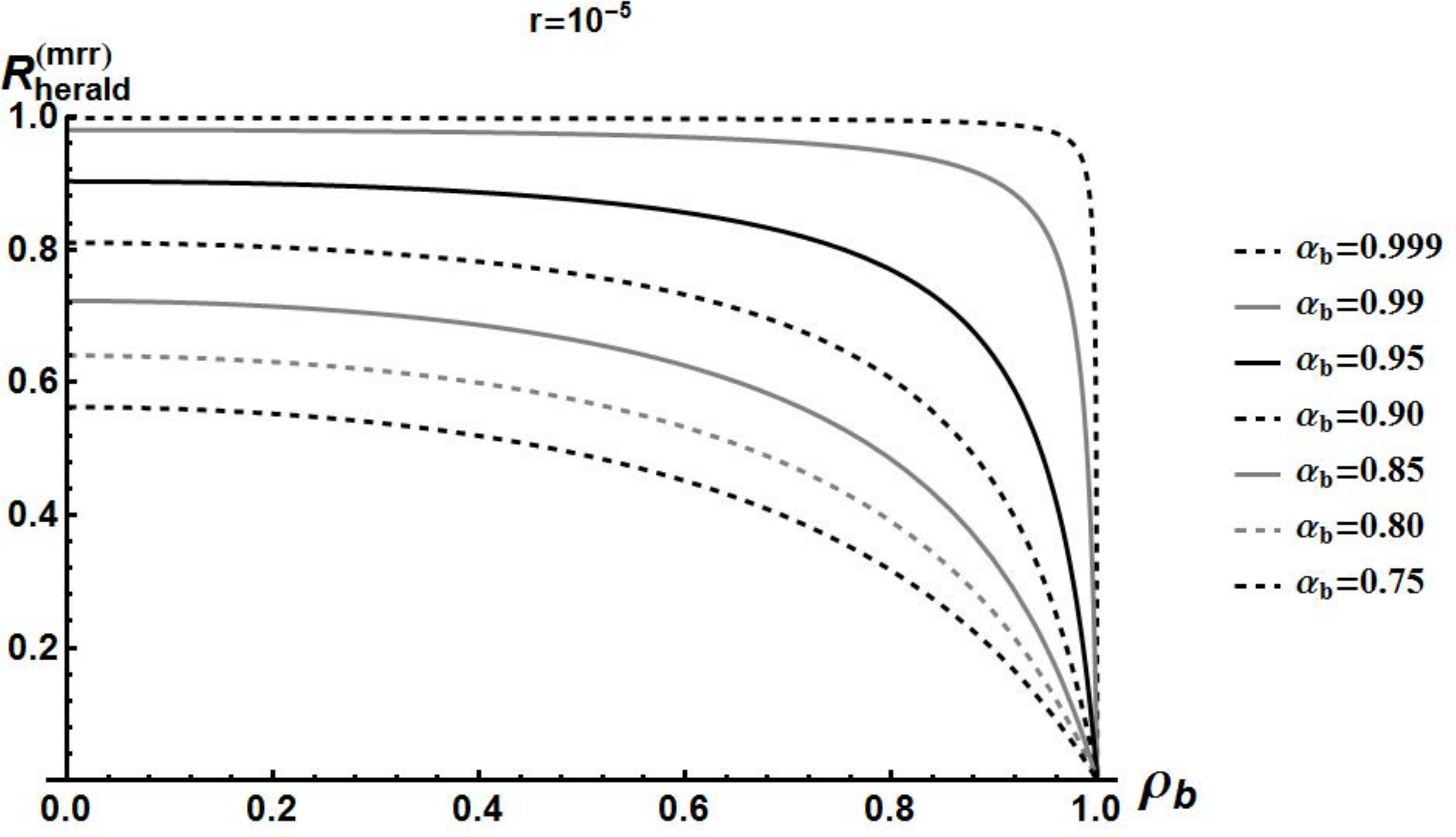} & %{fig6_ppg_left_Rheraldmrr_r1e-5_rho_alphas0p999-0p75_v3}
\includegraphics[width=3.5in,height=2.5in]{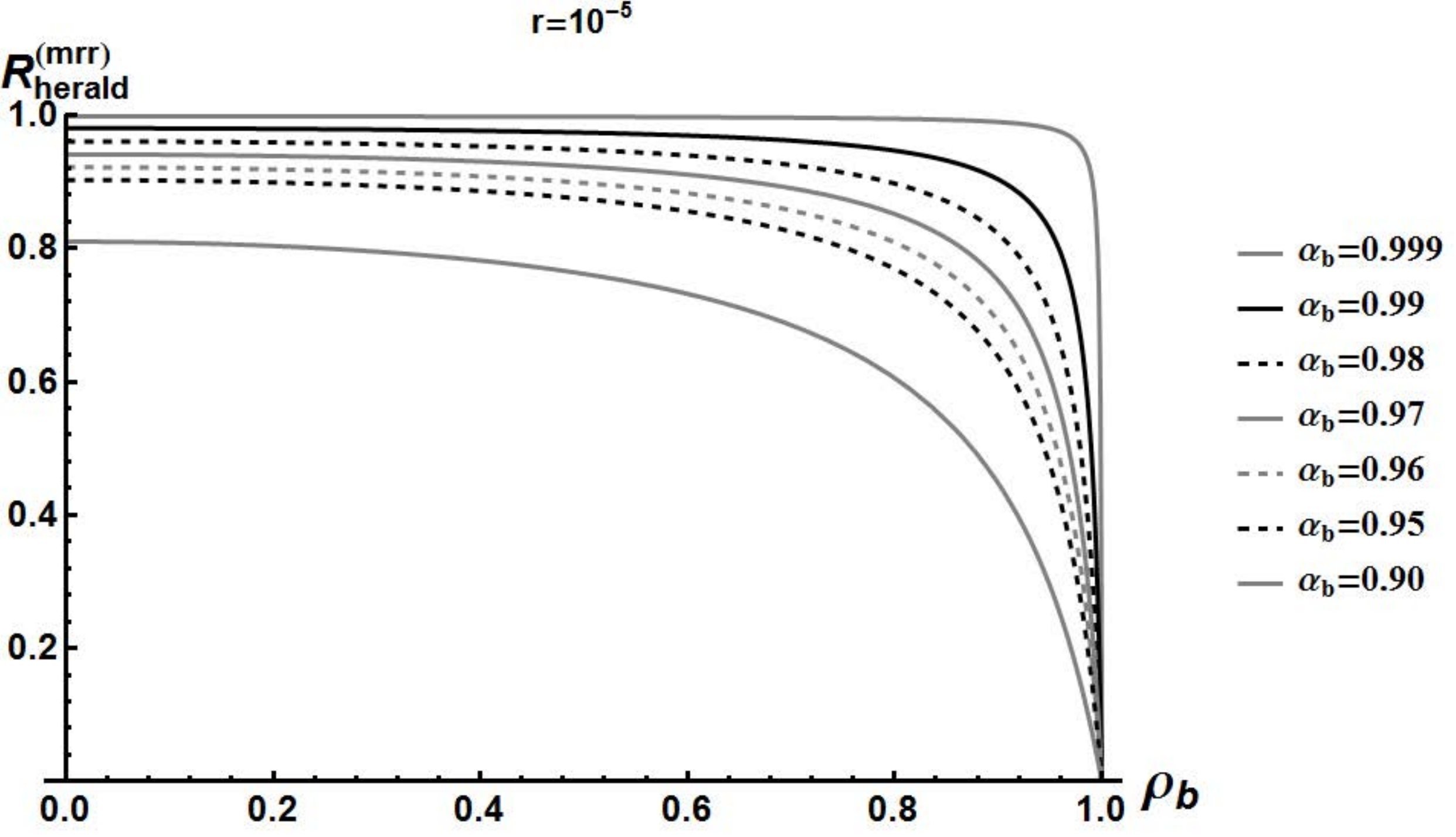}%{fig6_ppg_right_Rheraldmrr_r1e-5_rho_alphas0p999-0p90_v3}
\end{tabular}
\caption{Heralding efficiency
$R^{(mrr)}_{\textrm{herlad}}(\om)$ with
$r=10^{-5}$ for (left) $\alpha = (0.999, 0.99, 0.95, 0.90, 0.85, 0.80, 0.75)$.
and the operationally relevant values (right) $\alpha = (0.999, 0.99, 0.98, 0.97, 0.95, 0.95, 0.90)$.
}\label{Rherald:mrr:r1e-5:rho:alphas}
\end{figure}
%=======================================================
In \Fig{Rherald:mrr:r1e-5:rho:alphas} we plot $R^{(mrr)}_{\textrm{herald}}(\om)$ with
$r=10^{-5}$ for the internal propagation loss values
(left) $\alpha~=~(0.99, 0.95, 0.90,$ $ 0.85, 0.80, 0.75)$, and for the operationally relevant
(for $\alpha\le 0.99$) internal propagation loss values $\alpha = (0.999, 0.99, 0.98, 0.97, 0.95, 0.95, 0.90)$.
Even for  high values of loss ($\alpha\le 0.95$), the heralding efficiencies remain relatively high over a broad range of the coupling parameter $\rho_b$.

%======================================================
%\newpage
\section{Summary and Discussion}\label{sec:summary:discussion}
In this work we have investigated photon pair generation via SPDC and SFWM in a single bus microring resonator using a formalism that explicitly takes into account the round trip circulation of the fields inside the cavity. We investigated
the biphoton generation-, coincidence to accidental, and heralding efficiency rates as function of the
bus-mrr coupling loss $\rho = e^{-\gamma\,T\,/2}$ and internal propagation loss $\alpha=e^{-\gamma'\,T\,/2}$ at rates $\gamma$ and $\gamma'$, respectively (with $T$ the roundtrip circulation time of the field(s)).
We showed \Eq{aout:matrix:form} that the signal-idler output fields $\vec{a}_{out}(\om)$ can be expressed in terms of
the input fields $\vec{a}_{in}(\om)$ and quantum noise operators $\vec{f}(\om)$ as
$\vec{a}_{out}(\om) = G(\om) \, \vec{a}_{in}(\om) + H(\om)\, \vec{f}(\om)$. The matrix $G(\om)$ encodes the classical phenomenological loss (for $\alpha <1$) \cite{Yariv:2000,Rabus:2007} of the mrr, while the matrix $H(\om)$ incorporates the coupling and internal propagation loss due to the quantum Langevin  noise fields $\vec{f}(\om)$ required to preserve unitarity of the composite system (signal-idler) and environment (noise) structure. While the standard Langevin input-output formalism often used in the literature is valid in the high cavity Q limit
($\rho\approx 1-\gamma\,T/2\rightarrow 1,\,\om\,T\ll 1$),  and near cavity resonances, the formulation developed here is valid throughout the free spectral range of the mrr.
We explored values of the noise field commutators which were uniquely derived by invoking the unitarity of the input and output fields (which required the later's commutators to have the canonical form for free fields).
For unequal signal and idler group velocities the cross noise commutators were non-zero, while in general, the noise commutators contained pump dependent contributions.
%We further investigated the effect of coupling and internal propagation loss on the entanglement of the biphoton state generated within the mrr (given by the log negativity of the Gaussian biphoton state), and showed that increased entanglement is favored by the high cavity Q limit ($\rho\rightarrow 1$) and low internal propagation loss
%($\alpha\approx 1-\gamma'\,T/2\rightarrow 1$).

This work purposely concentrated on the weak (undepleted) pump limit and perfect phase matching in order to focus on the influence of the mrr coupling $\rho$ and internal propagation loss $\alpha$ parameters.
As indicated earlier in this work, non-zero phase matching can be straightforwardly included, which modifies the $G$ and $H$ matrices with multiplicative $\sin$c function contributions. Similarly, this work only included effects of dispersion through the mrr round trip times $T_k = L/v_k$ for $k\in\{a, b\}$ for the signal ($a$) and idler fields ($b$)
with possibly different group velocities $v_k$. Expansion of the frequency dependent momentum vectors for the signal and idler fields about a central frequency could also be straightforwardly accommodated.
A further logical extension of this work would be to consider the strong pump field regime in the spirit of the recent work by Vernon and Sipe \cite{Sipe:2015b} where effects such as pump depletion, and self-phase and cross-phase modulation could be taken into account.

\appendix*
\section{The high cavity Q limit}\label{app}

\subsection{The high Q limit and reduction to the standard Langevin input-output formalism for a single mrr field}\label{app:derivation:single:bus:highQ:limit}
Both Raymer and McKintrie \cite{Raymer:2013} and Alsing \textit{et al.} \cite{Alsing_Hach:2016} considered the
comparison of their formulations to the high Q limit.
Raymer and McKintrie define the high Q limit through the physical conditions (see \cite{Raymer:2013}, Section III)
(i) the cross coupling $\tau_a$ is very small so that the cavity storage time is long,
(ii) the cavity round trip time $T_a$ is small compared to the duration of the input-field pulse i.e. $\om\,T_a \ll 1$, and
(iii) the input field is narrow band and thus well contained within a single FSR of the mrr.
By defining (now including internal loss)
\bea{RM:HCQ:limit}
\rho_a & \equiv & e^{-\g_a T_a/2} \approx 1-\g_a T_a/2,
\qquad \tau_a = \sqrt{1-\rho_a^2} \approx  \sqrt{\g_a T_a}, \no
 \alpha_a &=& e^{-\g'_a T_a/2} \approx 1-\g'_a T_a/2 ,
\qquad  e^{i\,\om\,T_a} \approx 1 + i\,\om\,T_a,
\eea
one has from \Eq{aout:AH:v3}
\bsub
\bea{G:HCQ:limit}
G_{out,in}(\om) &=& \left(
\frac{\rho_a-\alpha_a\,e^{i\theta_a}}{1-\rho_a\,\alpha_a\,e^{i\theta_a}}
\right)
\quad \myover{{\longrightarrow} {\textrm{high Q}}} \quad \frac{i\om + (\g_a-\g'_a)/2}{-i\om + (\g_a+\g'_a)/2}, \\
H_{out,in}(\om) \equiv |H_{out,in}(\om)| &=& \frac{|\tau_a|^2\,(1-\alpha_a^2)}{|1-\rho_a\,\alpha_a\,e^{i\,\theta_a}|^2}
\;\quad \myover{{\longrightarrow} {\textrm{high Q}}} \quad \frac{\sqrt{\g_a\,\g'_a}}{\om^2 + [(\g_a+\g'_a)/2]^2},
\eea
\esub
where, without loss of generality, we have taken the phase of $H_{out,in}(\om)$ to be zero (or equivalently, absorbed into the definition of the noise operator $f_a(\om)$).

Under the assumptions of the high cavity Q limit one has $a(L_-,t)\approx a(0_+,t)$. Raymer and McKinstrie \cite{Raymer:2013} show that by defining the rescaled cavity field as $a(t) \equiv \sqrt{T_a}\,a(0_+,t)$
and considering the transfer function $G_{0_+,in}(t)$ in the time domain,
the equation of motion (without noise) becomes $\partial_t\,a(t) = -\half\g'_a\,a(t) + \sqrt{\g'_a}\,a_{int}(t)$.
Additionally, the output boundary condition \Eq{RM:3:v3}, in the limit $\rho_a\rightarrow 1$, $\tau_a\rightarrow \sqrt{\g'_a\,T_a}$ becomes
$a_{out}(t)= \sqrt{\g'_a}\,a(t) - a_{in}$, which is the standard Langevin boundary condition  %\Eq{eqn:Langevin:input:output:BC}
$a_{in} + a_{out} = \sqrt{\gamma_a}\,a$ \cite{Walls_Milburn:1994,Orszag:2000}.

\subsection{The high cavity Q limit of $G(\om)$ and $H(\om)$}\label{app:highQ:limit:G:H}
The high cavity Q limit is defined by (see Raymer and McKinstrie \cite{Raymer:2013})
$\rho_k\equiv e^{-\gamma_k\,T_k/2}\approx 1-\gamma_k\,T_k/2$ for $k\in\{a, b\}$ which implies
$\tau_k^2 \approx \gamma_k\,T_k$, and by taking the limit $\om\,T_k\ll 1$ so that $e^{i\theta_k}\approx 1 + i\,\om\,T_k$. If we further assume that the internal propagation loss is small we can also take $\alpha_k\approx 1-\gamma'_k\,T_k/2$.
We then have $S_k = (1-\rho_k\alpha_k\,e^{i\,\theta_k})^{-1}\approx [(s + \G_k/2)\,T_k]^{-1}$, a complex Lorentzian lineshape factor,
where for simplicity we have defined $s\equiv -i\om$
($s$ can be considered as a Laplace transform solution variable),
and have defined the total decay rate $\G_k = \gamma_k + \gamma'_k$.
Let us also further define $\Delta_k = \gamma_k - \gamma'_k$
as the difference between the coupling and internal propagation losses.
Then, $\exik -\rho_k \rightarrow (-s+\Delta_k/2)\,T_k$
and
%$\tilde{D}_{ab} \approx D(s)/[(s + \G_a/2)\,(s + \G_b/2)$ where
$D(s)\approx \left[(s + \G_a/2)\,(s + \G_b/2) - |g\alpha_p|^2\right]\,T_a\,T_b\equiv\tilde{D}(s)\,T_a\,T_b$.
We then obtain from \Eq{Gom:v2:matrix:elements}
\bea{G:HighQ:limit}
&{}& G(\om) \myover{{\longrightarrow} {\textrm{high Q}}} \no
&{}&
\frac{1}{\tilde{D}(s)} \,
\left(
  \begin{array}{cc}
    (-s + \Delta_a/2)\,(s+\G_b/2)+ |g\alpha_p|^2\,[1-\gamma_a\,T_a/2] & i\,g\,\alpha_p\,\sqrt{\gamma_a\,\gamma_b} \,\sqrt{T_b/T_a}\,[1-(s+\g'_b/2)\,T_b] \\
   -i\,g\,\alpha^*_p\,\sqrt{\gamma_a\,\gamma_b} \,\sqrt{T_a/T_b}\,[1-(s+\g'_a/2)\,T_a] &  (-s + \Delta_b/2)\,(s+\G_a/2)+ |g\alpha_p|^2\,[1-\gamma_b\,T_b/2],
  \end{array}
\right). \qquad\;\;
\eea

For the noise terms, let us redefine the noise operators as $\tilde{f}_k(\om) \equiv (1-\alpha_k^2)^{-1/2}\,f_k(\om)$ for $k\in\{a,b\}$
and equivalently the values of the commutators as
$[f_k(\om), f^\dag_k(\om)]=C_{kk}(\om) \equiv (1-\alpha_k^2)\,\tilde{C}_{kk}(\om)$ and
$D_{ab}(\om) \equiv (1-\alpha_a^2)^{1/2}\,(1-\alpha_b^2)^{1/2}\tilde{D}_{ab}(\om)$, so that
$[\tilde{f}_k(\om) , \tilde{f}_k(\om')] = \tilde{C}_{kk}(\om)\,\delta(\om-\om')$ and
$[\tilde{f}_a(\om) , \tilde{f}_b(\om')] = \tilde{D}_{ab}(\om)\,\delta(\om-\om')$.
Then
\be{aout:tilde:H}
\vec{a}_{out} \equiv G(\om)\, \vec{a}_{in} + \tilde{H}(\om)\,\vec{\tilde{f}}(\om),  \quad \tilde{H}(\om)\, = H(\om)\,\Lambda_{\alpha},
\quad  \Lambda_{\alpha} \equiv
\left(
  \begin{array}{cc}
   (1-\alpha_a^2)^{1/2}  & 0 \\
    0 & (1-\alpha_b^2)^{1/2} \\
  \end{array}
\right).
\ee
From \Eq{Hom:v2} and \Eq{aout:tilde:H} in the high Q limit, where $(1-\alpha^2_k)^{1/2}\rightarrow \sqrt{\g'_k\,T_k}$, we then have
\be{H:HighQ:limit}
\tilde{H}(\om) \myover{{\longrightarrow} {\textrm{high Q}}}
\frac{1}{\tilde{D}(s)} \,
\left(
  \begin{array}{cc}
    \sqrt{\gamma_a\,\gamma'_a}\,(s+\G_b/2) & i\,g\,\alpha_p\,\sqrt{\gamma_a\,\gamma'_b} \,\sqrt{T_b/T_a} \\
   -i\,g\,\alpha^*_p\,\sqrt{\gamma'_a\,\gamma_b} \,\sqrt{T_a/T_b} &  \sqrt{\gamma_b\,\gamma'_b}\,(s+\G_a/2),
  \end{array}
\right). \qquad\;\;
\ee
Except for the extra correction factors
%of $-|g\alpha_p|^2\,T_k/2$
indicated in the square brackets
in $G(\om)$ (which can be safely approximated as unity to lowest order in $|g\,\alpha_p|$)
these matrices are the same expressions as obtained by Tsang (see (4.11) in \cite{Tsang:2011})
using the standard Langevin input/output procedure and assuming $T_a=T_b=L/v$.

Note further that to zeroth order in $|g\alpha_p|$ we have
$D^{-1}\approx S_a\,S_b/(T_a\,T_b)\rightarrow [(s + \G_a/2)^{-1}\,(s + \G_b/2)\,T_a\,T_b]^{-1}$
and thus $G(\om)$ reduces in first order in $|g\alpha_p|$ to
\be{G:HighQ:limit:1storder}
% G(\om) \myover{{\longrightarrow} {\textrm{high Q}\, \mathcal{O}(|g\,\alpha_p|) }}
%
G(\om) \myover{ {\longrightarrow} {\stackrel{\textrm{\scriptsize high Q}}{\mathcal{O}(|g\,\alpha_p|)}} }
\left(
  \begin{array}{cc}
    \fracd{ -s + \Delta_a/2}{ s+\G_a/2}   & i\,g\,\alpha_p\,\left(\fracd{\sqrt{\gamma_a}}{s+\G_a/2}\right)\,\left(\fracd{\sqrt{ \gamma_b}}{s+\G_b/2}\right)  \\
    -i\,g\,\alpha^*_p\,\left(\fracd{\sqrt{\gamma_a}}{s+\G_a/2}\right)\,\left(\fracd{\sqrt{\gamma_b}}{s+\G_b/2}\right)  &  \fracd{-s + \Delta_b/2}{s+\G_b/2},
  \end{array}
\right), \qquad\;\;
\ee
where we have also used $\exik\approx 1$.
In this limit, the diagonal terms $G_{kk}$, which directly couple $(\vec{a}_{out})_k$ to $(\vec{a}_{in})_k$ for $k\in\{a, b\}$, have same frequency dependent shifts of the output signal-idler fields relative to the internal signal-idler fields as given by the conventional Langevin approach \cite{Haus:1984,Walls_Milburn:1994,Orszag:2000}.
The lower order (in $|g\,\alpha_p|$) off diagonal terms $G_{ab}$ and $G_{ba}$ contain the product of
%$\sqrt{\gamma_a}/(s+\G_a/2)$ $*\sqrt{\gamma_b}/(s+\G_a/2)$ which are the product of the
Lorentzian lineshape factors $\sqrt{\gamma_k}/(s+\G_k/2)$  relating the output signal-idler fields  to the opposite idler/signal fields inside the cavity.
Similarly, for $\tilde{H}(\om)$ we have
\be{H:HighQ:limit:1storder}
% G(\om) \myover{{\longrightarrow} {\textrm{high Q}\, \mathcal{O}(|g\,\alpha_p|) }}
%
\tilde{H}(\om) \myover{ {\longrightarrow} {\stackrel{\textrm{\scriptsize high Q}}{\mathcal{O}(|g\,\alpha_p|)}} }
\left(
  \begin{array}{cc}
   \fracd{\sqrt{\g_a\,\g'_a}}{s+\G_a/2}   & \fracd{i\,g\,\alpha_p\,\sqrt{\g_a\,\g'_b}}{(s+\G_a/2)\,(s+\G_b/2)}  \\
    \fracd{i\,g\,\alpha^*_p\,\sqrt{\g'_a\,\g_b}}{(s+\G_a/2)\,(s+\G_b/2)}  & \fracd{\sqrt{\g_b\,\g'_b}}{s+\G_b/2}
  \end{array}
\right).
\ee

\subsection{Biphoton generation rate $R_{ab}(\om)$ in the high $Q$ limit}\label{app:highQ:limit:Rab}
To make connection with other works, let us more closely examine the two-photon generation rate  given by
$R_{ab}(\om) = |\rab|^2\,|\psi^{(2)}_{ab}(\om)|^2$ in the high cavity Q limit.
Note that from \Eq{defns:S_D_dab:v2} we can write the $D(s)$ in \Eq{GLom} and \Eq{HLom} as
$D(s) = (1 - \rho_a\,\alpha_a\,e^{-s\,T_a})\,(1 - \rho_b\,\alpha_b\,e^{-s\,T_b}) - |g\,\alpha_p|^2\,T_a\,T_b$ where $s=-i\,\om$.
The pole structure of $D(s)$ is obtained by the roots $s_{\pm}$ of $D(s_{\pm})=0$. In general this is a transcendental equation which must be solved numerically.
If we approximate $e^{-s\,T_k}\approx 1 -s\,T_k$, we obtain a quadratic equation  in $s$ with poles $s_\pm$, and
\bsub
\bea{D:roots}
D(s) &\approx& (s-s_+)\,(s-s_-),   \no
s_{\pm} &=& \half\,\left( \fracd{y_a}{x_a} + \fracd{y_b}{x_b} \right)
\pm \sqrt{ \left[\half\,\left( \fracd{y_a}{x_a} - \fracd{y_b}{x_b} \right)\right]^2 + |g\,\alpha_p|^2 }, \quad
x_k = \rho_k\,\alpha_k, \quad y_k = \fracd{(1-\rho_k\,\alpha_k)}{T_k},\qquad\quad \label{D:roots:exact} \\
&\myover{{\longrightarrow} {\textrm{high Q}}}&
\left( \fracd{\G_a + \G_b}{4} \right)
\pm \sqrt{ \left(\fracd{\G_a - \G_b}{4}\right)^2 + |g\,\alpha_p|^2 } \; \equiv \;  \pi_{\pm},
\\
&\myover{ {\longrightarrow} {\stackrel{\textrm{\scriptsize high Q}}{\mathcal{O}(|g\,\alpha_p|)}} } &
\left\{
\begin{array}{c}
  -\G_a/2 \\
  -\G_b/2
\end{array}
\right. \quad \textrm{for} \quad \G_b > \G_a, \label{D:roots:exact:highQ}
\eea
\esub
where $\pi_{\pm}$ are the poles of $D(s)$ as computed by Tsang \cite{Tsang:2011} using a standard Langevin input-output calculation.
Then, the two-photon generation rate becomes to lowest order in $|g\,\alpha_p|^2$
\bsub
\bea{Rab:L}
R_{ab}(\om) &=& |\rab|^2\,|\psi^{(2)}_{ab}(\om)|^2 \approx |\rab|^2\,|G^{(L)*}_{aa}(\om)\,G^{(L)}_{bb}(\om)|^2, \no
& \myover{ {\longrightarrow} {D(s)\approx (s-s_+)\,(s-s_-)} } &   % D(s)\approx (s-s_+)\,(s-s_-) % e^{-s\,T_k}\approx 1-s\,T_k
|g\,\alpha_p|^2 \,
 \fracd{ \left(\frac{\tau_a}{\sqrt{T_a}}\right)^2\,\left|\frac{1-\rho_b\,\exib}{T_b}\right|^2\,\left| \exia \right|^2}{(\om^2 + s^2_+)\,(\om^2 + s^2_-)}  \cdot
 \fracd{ \left(\frac{\tau_b}{\sqrt{T_b}}\right)^2\,\left|\frac{1-\rho_a\,\exib}{T_a}\right|^2\,\left| \exib \right|^2}{(\om^2 + s^2_+)\,(\om^2 + s^2_-)}, \qquad\;\;\;  \label{Rab:L:pma}\\
&\myover{{\longrightarrow} {\textrm{high Q}}}&
|g\,\alpha_p|^2 \,
 \fracd{\g_a\,\left[\om^2 + (\G_b/2)^2\right]}{(\om^2 + s^2_+)\,(\om^2 + s^2_-)}  \cdot
 \fracd{\g_b\,\left[\om^2 + (\G_a/2)^2\right]}{(\om^2 + s^2_+)\,(\om^2 + s^2_-)}, \label{Rab:L:pma:Tsang}\\
& \myover{ {\longrightarrow} {\stackrel{\textrm{\scriptsize high Q}}{\mathcal{O}(|g\,\alpha_p|^2)}} } &
 |g\,\alpha_p|^2\,\fracd{\g_a}{[\om^2 + (\G_a/2)^2]}\cdot \fracd{\g_b}{[\om^2 + (\G_b/2)^2]}, \label{Rab:L:pma:Scholz}
%
%& \myover{ {\longrightarrow} {\stackrel{\textrm{\scriptsize high Q}}{\mathcal{O}(|g\,\alpha_p|^2)}} } &
% \fracd{|g\,\alpha_p|^2 \,\g_a\,\g_b\,}{[\om^2 + (\G_a/2)^2]\,[\om^2 + (\G_b/2)^2]},
\eea
\esub
where in the third line we have used $|\exik|^2\approx 1$ and in the fourth line we have used \Eq{D:roots:exact:highQ}.
The above expressions generalize two photon rate $R_{ab}(\om) = |g\,\alpha_p|^2 \,\g_a\,\g_b/[(\om^2 + s^2_+)\,(\om^2 + s^2_-)]$
computed by Tsang \cite{Tsang:2011}, which to $\mathcal{O}(|g\,\alpha_p|^2)$ agrees with \Eq{Rab:L:pma:Tsang}.
The last line \Eq{Rab:L:pma:Scholz} is the form computed by Scholz using the (complex) Lorenztian modified form
$\sqrt{\gamma_a}\,a(\om)/(s+\G_a/2)$ and $\sqrt{\gamma_b}\,b^\dag(\om)/(-s+\G_b/2)$ for the field operators inside the mrr.
The expression \Eq{Rab:L:pma}, quadratic in the poles $s_\pm$, more fully takes into account the effect of the
the field circulation factors $S_k = 1/[1-\rho_k\,\exib]$ on the two-photon generation rate.

%======================================================
%\newpage
\begin{acknowledgments}
PMA, would like to acknowledge support of this work from
Office of the Secretary of Defense (OSD) ARAP QSEP program, and thank
J. Schneeloch and M. Fanto for helpful discussions.
EEH would like to acknowledge support for this work was provided by the Air Force Research
Laboratory (AFRL) Visiting Faculty Research Program
(VFRP) SUNY-IT Grant No. FA8750-13-2-0115.
%Some of the authors were also supported by the Air Force Office
%of Scientific Research under Grant No. FA9550-10-1-0217.
Any opinions, findings and conclusions or recommendations
expressed in this material are those of the author(s) and do not
necessarily reflect the views of Air Force Research Laboratory.
\end{acknowledgments}
%============================

%--------------------------------------------
%\newpage
\bibliography{rr_losses_refs}
%-------------------------------------------
%\begin{thebibliography}{99}
%%
%\bibitem{Wei_Norman:1963} J. Wie and E. Norman "Lie algebraic soluition of linear differential equations," J. Math. Phys. {\textbf{4}}, 575  (1963).
%%
%\bibitem{Agarwal:2013} G.S. Agarwal, {\textit{Quantum Optics}}, Cambridge Univ. Press (2013).
%%
%\end{thebibliography}
%--------------------
\end{document}